\documentclass{aa}
    \usepackage{graphicx}

    \usepackage[dvipsnames,svgnames,x11names]{xcolor}

    \usepackage{amsmath}
    \usepackage{amssymb}
    \usepackage{txfonts}
    \usepackage{adjustbox}
    \graphicspath{{./figures/}}

    \usepackage{natbib}
    \bibpunct{(}{)}{;}{a}{}{,} % to follow the A&A style

    \newcommand{\unsim}{\mathord{\sim}}
    \newcommand{\myendash}{\ensuremath{\,\text{--}\,}}

    \newcommand{\gas}{\ensuremath{_\mathrm{gas}}}
    \newcommand{\sph}{\ensuremath{_\mathrm{SPH}}}
    \newcommand{\amuse}{\textsc{amuse}}
    \newcommand{\bridge}{\textsc{bridge}}
    \newcommand{\ficode}{\textsc{fi}}
    \newcommand{\gadget}{\textsc{gadget2}}
    \newcommand{\fastkick}{\textsc{fastkick}}
    \newcommand{\huayno}{\textsc{huayno}}
    \newcommand{\seba}{\textsc{SeBa}}
    \newcommand{\stellarwind}{\mbox{\textsc{stellar\_wind.py}}}

    \newcommand{\MSun}{\ensuremath{\mathrm{M}_\odot}}
    \newcommand{\RSun}{\ensuremath{\mathrm{R}_\odot}}
    \newcommand{\LSun}{\ensuremath{\mathrm{L}_\odot}}
    \newcommand{\Mdot}{\ensuremath{\dot{M}}}
    \newcommand{\vinf}{\ensuremath{v_{\infty}}}
    \newcommand{\vinit}{\ensuremath{v_{0}}}
    \newcommand{\vesc}{\ensuremath{v_{\mathrm{esc}}}}
    \newcommand{\vphesc}{\ensuremath{v_{\mathrm{phesc}}}}
    \newcommand{\rmax}{\ensuremath{r_{\mathrm{max}}}}
    \newcommand{\raccstart}{\ensuremath{r_{\mathrm{acc\_start}}}}
    \newcommand{\rmid}{\ensuremath{r_{\mathrm{mid}}}}
    \newcommand{\Emech}{\ensuremath{E_{\mathrm{mech}}}}
    \newcommand{\Lmech}{\ensuremath{L_{\mathrm{mech}}}}
    \newcommand{\tcrit}{\ensuremath{t_{\mathrm{c}}}}
    \newcommand{\tcode}{\ensuremath{t_{\mathrm{code}}}}
    \newcommand{\ttot}{\ensuremath{t_{\mathrm{tot}}}}
    \newcommand{\tnbody}{\ensuremath{t_{\mathrm{N-body}}}}

    \newcommand\T{\rule{0pt}{2.6ex}}
    \newcommand\B{\rule[-1.2ex]{0pt}{0pt}}

    \newcommand{\inprep}[1]{(#1 in prep.)}

\begin{document}
    \title{Simulating stellar winds in AMUSE}

    % \author[E. van der Helm et al.]{
    \author{
        Edwin van der Helm \inst{1} \and
        Martha I. Saladino \inst{2,1} \and
        Simon Portegies Zwart \inst{1} \and
        Onno Pols \inst{2}
        }

    \institute{
        Leiden Observatory, Leiden University, PO Box 9513, 2300 RA, Leiden, The Netherlands \and
        Department of Astrophysics/IMAPP, Radboud University, PO Box 9010, 6500 GL Nijmegen, The Netherlands
        }

    \date{Received October 2, 2017; Accepted March 23, 2019}

    \abstract{}{
        We present \stellarwind, a module that provides multiple methods of simulating stellar winds using smoothed particle hydrodynamics codes (SPH) within the astrophysical multipurpose software environment (\amuse) framework.
        }{
        The module currently includes three ways of simulating stellar winds:
        With the simple wind mode, we create SPH wind particles in a spherically symmetric shell for which the inner boundary is located at the radius of the star.
        We inject the wind particles with a velocity equal to their terminal velocity.
        The accelerating wind mode is similar, but with this method particles can be injected with a lower initial velocity than the terminal velocity and they are accelerated away from the star according to an acceleration function.
        With the heating wind mode, SPH particles are created with zero initial velocity with respect to the star, but instead wind particles are given an internal energy based on the integrated mechanical luminosity of the star.
        This mode is designed to be used on longer timescales and larger spatial scales compared to the other two modes and assumes that the star is embedded in a gas cloud.
        }{
        We present a number of tests and compare the results and performance of the different methods.
        For fast winds, we find that both the simple and accelerating mode can reproduce the desired velocity, density and temperature profiles.
        For slow winds, the simple wind mode is insufficient due to dominant hydrodynamical effects that change the wind velocities.
        The accelerating mode, with additional options to account for these hydrodynamical effects, can still reproduce the desired wind profiles.
        We test the heating mode by simulating both a normal wind and a supernova explosion of a single star in a uniform density medium.
        The stellar wind simulation results matches the analytical solution for an expanding wind bubble.
        The supernova simulation gives qualitatively correct results, but the simulated bubble expands faster than the analytical solution predicts.
        We conclude with an example of a triple star system which includes the colliding winds of all three stars.
        }{}

    \maketitle

\section{Introduction}
        % Stars have wind, which is interesting
        Stars lose mass through stellar winds during various stages of their evolution \citep[e.g.][]{meyer-vernet_basics_2007,owocki_stellar_2013,puls_physics_2015}.
        These winds can affect the gas near the star, creating lower density bubbles \citep{castor_interstellar_1975} and regulating star formation \citep{oey_massive_2009}.
        If a binary companion is present, accretion of the stellar wind material can also affect the evolution of that companion \citep{boffin_mass_2015}.

        % There are two important parameters and three types of wind
        The two most important parameters of the stellar wind are the mass-loss rate, \Mdot, and the terminal wind velocity, \vinf, which determine the effect of the wind on the environment.
        Based on these parameters, stellar winds can be broadly divided into three categories \citep{owocki_stellar_2013}:
        1) Cool main-sequence stars like the Sun have winds with very low mass-loss rates ($\Mdot \unsim 10^{-14}$~\MSun/yr) that are thermally or gas pressure driven.
        2) Cool giants and super giants have slow ($\vinf \unsim 5 \myendash 30$~km/s) high mass-loss rate ($\Mdot \unsim 10^{-7} \myendash 10^{-5}$~\MSun/yr) winds driven mainly by radiation pressure on dust \citep{hofner_wind_2015}.
        3) Hot luminous stars have fast ($\vinf \unsim 200 \myendash 3000$~km/s) high mass-loss rate ($\Mdot \unsim 10^{-7} \myendash 10^{-5}$~\MSun/yr) line driven winds \citep{puls_mass_2009}.
        The second and third category have the highest kinetic output and therefore have the most pronounced effect on the stellar environment (not including cumulative effects).

        % Stellar wind has been simulated in many ways
        To simulate stellar winds in detail, a combination of hydrodynamics, radiative transfer, dust formation and chemical abundances is required.
        Such simulations have been done for many years although they are extremely computationally expensive.
        In most cases simulations are limited to 1D or 2D models \citep[e.g.][]{owocki_time-dependent_1988,blondin_hydrodynamic_1990,kudritzki_winds_2000,boffin_mass_2015}.
        To investigate the net effect of the stellar wind on the environment, it is often sufficient to simulate the stellar wind using only hydrodynamics    \citep{theuns_wind_1993,cuadra_galactic_2006,mohamed_3d_2012}.
        For larger scale simulations, stellar wind feedback is often included using a sub-grid model as it can influence star formation and launch galactic winds \citep[e.g.][]{agertz_toward_2013,muratov_gusty_2015}.

        % Amuse is great for coupled simulations and has been used for wind
        For all these simulations, the astrophysical multipurpose software environment \citep[\amuse\footnote{amusecode.org};][]{portegies_zwart_multi-physics_2013,pelupessy_astrophysical_2013,van_elteren_multi-scale_2014,AMUSE_book,AMUSE_zenodo} can be useful.
        It provides a uniform interface for many types of simulations with a large and growing set of simulation codes.
        The consistent python interface makes it possible to quickly set up a scientific simulation and easily exchange different parts of these simulations.
        While stellar winds have been simulated before using \amuse\ \citep[e.g.][]{pelupessy_evolution_2012}, a consistent and properly tested module was still missing.

        % We have created a wind module that makes it easy to simulate, in this paper ...
        The \stellarwind\ code presented in this paper can be combined with the SPH (Smoothed particle hydrodynamics), N-body, stellar evolution and (with some additional work) radiative transfer codes that are already available.
        We describe the \stellarwind\ code and explain the different modes in which it can be used (Section~\ref{sec: Methods}).
        In Section~\ref{sec: Tests} we present a series of tests in which we compare the results from the different modes with theoretical expectations and previous work.
        We conclude in Section~\ref{sec: Discussion and Conclusion} with an exposition of some ongoing research projects using this code and ideas for further use.

\section{Methods}
        \label{sec: Methods}

        % We have stars and create SPH particles around that
        The goal of \stellarwind\ is to create gas particles that represent the stellar wind from one or more stars.
        The code requires a number of stars, represented by \amuse\ particles\footnote{A particle set is the fundamental data structure in \amuse. It is an array of particles (stars, SPH particles etc) which contain information to control the data. Each element (particle) of the particle set has certain attributes, such as mass, position, velocity, etc.}, with stellar properties that can be derived from observations or stellar evolution simulations.
        Using this, SPH particles are created with the appropriate wind properties in an initially spherically symmetric shell with inner boundary at the radius of the star.
        The number of SPH particles is computed according to the mass-loss rate associated with the star undergoing mass loss and the predefined SPH particle mass, $M_\mathrm{SPH}$.
        These particles can be added to any SPH code in \amuse\ which simulates the hydrodynamics of the wind.

        % SPH codes, N-body codes and Bridge can be used for the rest
        Creating the SPH particles is only one step in the simulations for which \stellarwind\ is used.
        Following the goal of the \amuse\ framework, the other parts of the simulations are handled by specialized interchangeable codes.
        For the hydrodynamics, SPH codes such as \ficode\ \citep{pelupessy_numerical_2005} and \gadget\ \citep{springel_cosmological_2005} can be used.
        In many applications, the stars move, for which a large number of N-body codes are available.
        To couple the stellar dynamics to the hydrodynamics gravitationally, \bridge\ \citep{fujii_bridge:_2007} is available.
        The stellar properties on which the wind is based will generally be calculated using a stellar evolution code.
        Both parametrized \citep[e.g.][]{hurley_comprehensive_2000} and Henyey type \citep[e.g.][]{paxton_modules_2011} stellar evolution codes are available in \amuse.
        Any or all of these codes can be combined with \stellarwind\ to set up a wide variety of simulations (see Section \ref{sec: Tests}). For more information about the codes available within \amuse\ and examples of how to couple them, we refer the reader to \citet{AMUSE_book}.

    \subsection{Calculating stellar wind properties} \label{sec: wind from stev}
        % The main stellar properties are ..., which can be given directly or derived from stellar evolution
        To simulate the stellar winds, the stellar parameters (mass, radius, temperature and position) and wind parameters (mass-loss rate, initial and terminal wind velocity) are required.
        All these parameters can simply be set directly, however some of them can be derived directly from stellar evolution codes available in \amuse.

        The \stellarwind\ module includes user-friendly routines to derive some of the stellar parameters such as stellar mass, mass-loss rate, stellar radius and effective temperature from one of the stellar evolution codes within \amuse.
        However, the terminal wind velocity, \vinf, is not calculated by any code currently in \amuse.
        Determining \vinf\ requires detailed and computationally expensive stellar wind simulations which include radiative transfer.
        For this reason, in order to compute the terminal velocities of hot stars, we provide within \stellarwind\ a formula that has been fitted to observations of hot stars \citep{kudritzki_winds_2000}, and which, they claim, is valid for these stars within 20\%.
        The terminal velocity of the wind is given by:
        \begin{equation}
            \begin{split}
            \vinf & = C(T_*) \vphesc, \\
            \mathrm{where,} \\
            C(T_{*}) & =
            \begin{cases}
                1 & T_* \leq 10\,000~\mathrm{K}, \\
                1.4 & 10\,000~\mathrm{K} < T_* < 21\,000~\mathrm{K}, \\
                2.65 & T_* \geq 21\,000~\mathrm{K}, \\
                \end{cases} \\
            \vphesc & = \sqrt{2 g_* R_* \left ( 1 - \Gamma \right )}, \\
            g_* & = \frac{G M_*}{R_*^2}, \\
            \Gamma & = 7.66 \cdot 10^{-5} \sigma_e \frac{L_* / \LSun}{M_* / \MSun}, \\
            \sigma_e & = 0.398 \frac{ 1 + I_{\mathrm{He}} \mathrm{Y}}{1 + 4 \mathrm{Y}}, \\
            \end{split}
            \end{equation}
        where \vphesc\ is the photospheric escape velocity (similar to the escape velocity \vesc\ with a correction term for Thomson scattering), $G$ is the gravitational constant, $M_*$, $R_*$, $L_*$ and $T_*$ are the mass, radius, luminosity and effective temperature of the star respectively, $\Gamma$ is the ratio of radiative Thomson acceleration to gravitational acceleration, $\sigma_e$ is the Thomson absorption coefficient, $Y$ is the Helium fraction and $I_{\mathrm{He}}$ is the number of electrons per Helium nucleus (in this paper we use default values of $I_{\mathrm{He}} = 2$ and Y = 0.25).
        For cooler stars, $\vinf \approx \vphesc$ and this formula is still applicable (Kudritzki, private communication).

    \subsection{Simple wind}
        \label{sec: simple}
        % Create the wind with the final radial velocity (or compensate gravity)
        Within \stellarwind, there are currently three wind modes available.
        The simplest mode creates a spherical shell of particles around the star with radial velocity, $v(r) = \vinf$ and initial temperature equal to the effective temperature of the star.
        While this may sound simplistic, a similar setup has been used effectively for a number of scientific problems \citep[e.g.][]{mohamed_3d_2012} and it serves as a starting point for the two other modes described in Sections \ref{sec: accelerating}  and \ref{sec: heating}.
        When the gravitational attraction of the star on the wind is included in the simulation, however, this will not result in the desired terminal wind velocity.
        We therefore release the wind with a larger velocity $v(r) = \sqrt{{\vinf}^2 + \vesc(r)^2}$ where $\vesc(r) = \sqrt{2 G M_* / r}$ is the local escape velocity at the initial particle radius, $r$.
        We calculate this new velocity for each particle because $\vesc(r)$ can vary within the thin shell in which we create the particles.

        % Outer radius depends on dt and density follows the desired profile
        We set the outer radius (\rmax) of the shell of new particles at the radius that the innermost part of the previously released shell ($R_*$) should have reached in the elapsed simulation time $\delta t$ (see Appendix~\ref{sec: radius at time}).
        We scale the particle positions within the shell to follow the density profile matching the velocity profile as described in Appendix~\ref{sec: new particle radii}.

    \subsubsection{SPH and initial distributions}
        \label{sec: SPH}
        \begin{figure}[!tbp]
            \includegraphics[width=0.5\textwidth]{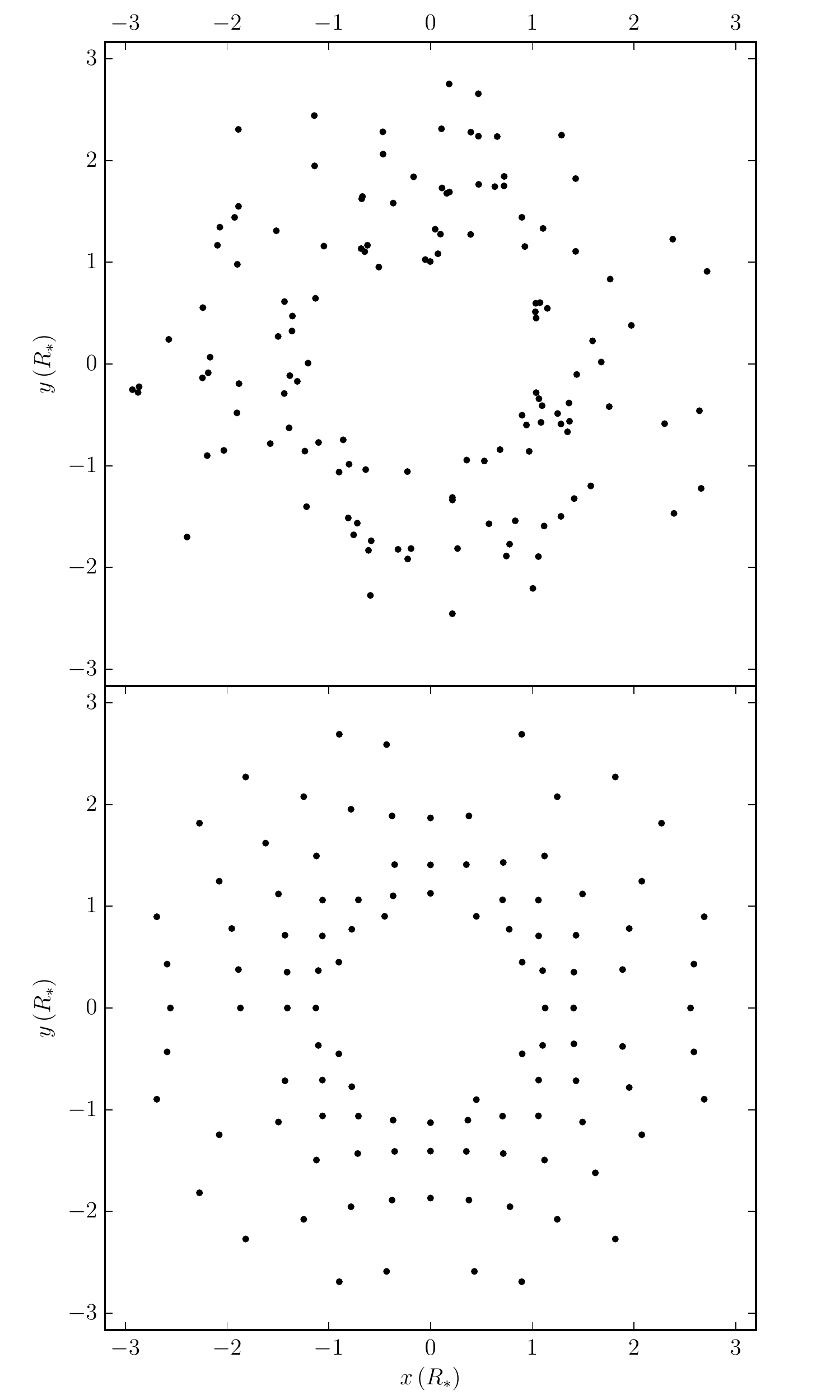}
            \caption{
                An example of the initial positions of newly created particles using a random (top) and grid (bottom) distribution.
                A shell of particles was created between 1 and 3 stellar radii ($R_*$) and the x and y positions of a thin slice ($|z| < 0.05\,R_*$) are shown.
                The positions are scaled to match the given density profile (see Appendix~\ref{sec: equations}).
                Note that this is merely an illustration of the difference between random and grid initial distributions.
                In real simulations, the shells would generally be much thinner.
                \label{fig: part positions}}
            \end{figure}

        % intro to SPH
        SPH is a method to solve the dynamics of a fluid by approximating it with a set of discrete particles \citep{monaghan_smoothed_1992}.
        Each particle has both a mass and a density, where the density is calculated using the distance to, and mass of, other particles that are nearby.
        To determine which other particles are taken into account (how nearby they have to be) a kernel function\footnote{In the SPH code used in this paper we use the spline kernel.} and a particle smoothing length ($h$) are used.
        In all modes of \stellarwind, the SPH particle mass is required to be fixed and the same for each particle.
        The smoothing length, $h$, is set adaptively by fixing the number of neighbouring particles that fall within one $h$ \citep[e.g. see][]{pelupessy_numerical_2005}.

        % Initial SPH positions are not trivial
        Creating an initial distribution of SPH particle positions is not trivial \citep[e.g.][]{diehl_generating_2015}.
        Randomly distributed positions are clumpy which can introduce shot noise that can affect the entire simulation.
        A better alternative is to have more regular spaces between particles, for instance a distribution that follows a grid.
        However, a regular grid tends to introduce preferred directions in the simulation that can affect the overall results.
        To solve this, it is common to start with either a random or grid distribution and let the system evolve (relax) to a steady state where the positions are regularly spaced without preferred directions \citep[for example a `glass' initial condition,][]{white_formation_1996,wang_discreteness_2007}.
        While some form of relaxation is preferred for simulations where all particles are created at once, for continuous particle creation like we describe here, this is not generally required.

        % Initial distribution can be random or (rotated) grid
        In \stellarwind\ we implement two methods in which wind particles can be initially distributed.
        One is a random distribution and the second follows a uniform grid.
        We present an example of both in Figure~\ref{fig: part positions}.
        The random initial distribution (top panel) is available so that users can investigate if it has advantages for their simulations.
        In this case, a shell with uniform density is created and then the radii are scaled to ensure the correct density profile.
        In the other option we have included (bottom panel), each new shell is created by cutting it out from a cube with positions following a uniform grid.
        The number of particles in this shell is generally not exactly the desired number of particles, $N_\mathrm{desired} = \delta t \cdot \dot{M}/M_\mathrm{SPH}$.
        We therefore remove a number of randomly selected particles from the grid (typically $\unsim 30\%$) to ensure the correct number of SPH particles.
        The grid can be randomly rotated each time a new shell is generated to avoid introducing preferred directions into the resulting wind.
        The positions of the particles are also radially scaled to ensure that the desired density profile is achieved.
        This is the cause of the curved appearance in the grid in Figure~\ref{fig: part positions}.

        % Why we don't use glass or WVT
        There are many more ways to create initial particle distributions.
        A good overview of the different methods and their advantages can be found in \citet{diehl_generating_2015}.
        Our method is a mix between the `stretched lattice' and the `shell' methods described there.
        The reason we do not use the more advanced methods described there is that they would require some form of computationally expensive relaxation for every new shell.
        This is a common issue with continuous particle creation methods.
        If the current methods are found to be unsatisfactory for a specific simulation, the code is set up in a modular way so adding a new particle distribution method is relatively easy.
        The uniform grid with random rotation is the default option used throughout this paper.
        However, due to the small number of particles in a single shell, the difference between this option and a random distribution is negligible for all the tests in Section~\ref{sec: Tests}.

    \subsection{Accelerating wind}
        \label{sec: accelerating}

        \begin{figure}[!tbp]
            \includegraphics[width=0.5\textwidth]{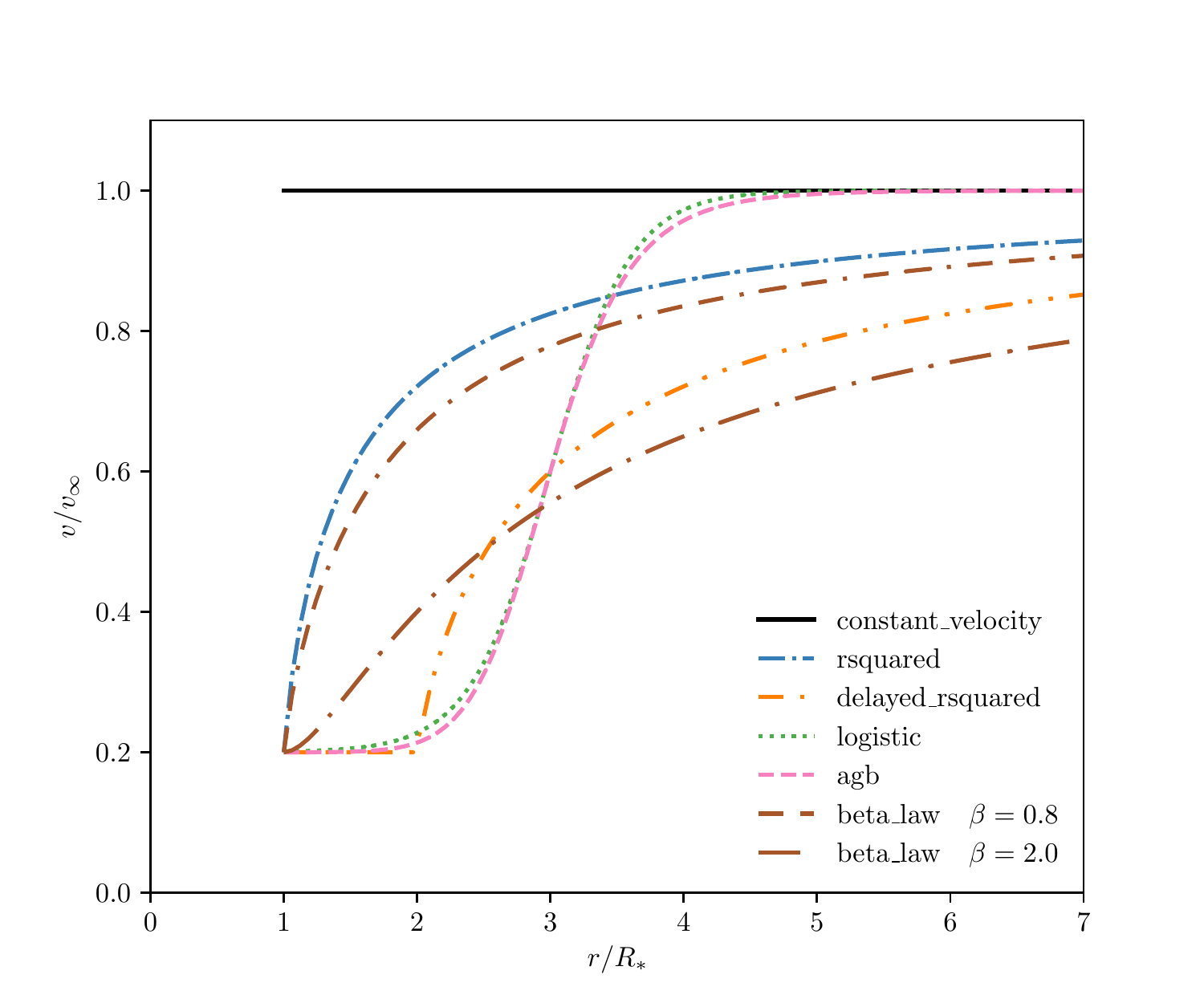}
            \caption{
                Radial velocity profiles for the wind acceleration functions currently available in \stellarwind.
                The formulae for these functions can be found in Table~\ref{tab: acc functions}.
                For the beta\_law we show two curves that are good fits for hot massive stars ($\beta = 0.8$) and cool giants ($\beta = 2.0$) \citep{lamers_introduction_1999}.
                To illustrate the shape of the acceleration curves, we have chosen $\vinit = 0.2 \, \vinf$,  $\raccstart = 2 \, R_*$ (for the delayed\_rsquared function), and $\rmid = 3 \, R_*$ and $s = \alpha = 10$ (for the logistic and agb function).
                \label{fig: acc functions}}
            \end{figure}

        \begin{table}
            \caption{
                An overview of the acceleration functions currently available in \stellarwind.
                Either the acceleration ($a$) or the velocity ($v$) is given depending on which is simpler.
                The corresponding $v$ or $a$ function can be derived using $a(r) = v(r) \frac{dv}{dr}$ and the known boundary conditions.
                Some functions allow user defined parameters to affect the functions (e.g. \raccstart, \rmid, $\beta$, etc.)
                The first three functions are rough approximations to the last three functions. Their advantage is that they are computationally faster.
                \label{tab: acc functions}}
		\begin{adjustbox}{width=0.49\textwidth}
            \begin{tabular}{l l l}
                \B Name & Equation & Use\\
                \hline
                \T constant\_velocity & $ v(r) = \vinf$ & Wide \\ [1mm]
                   & & binaries\\
                rsquared & \(\displaystyle a(r) \propto \frac{1}{r^2} \) & Hot stars \\ [4mm]
                delayed\_rsquared & \(\displaystyle a(r) \propto
                    \begin{cases}
                        0 & r < \raccstart \\
                        \frac{1}{r^2} & r \ge \raccstart  \\
                        \end{cases}
                    \) & Cool stars\\ [5mm]
                logistic & \(\displaystyle v(r) = \vinit + \frac{\vinf - \vinit}{1 + e^{-s \frac{r - \rmid}{\rmid}}} \)  & AGB winds\\ [5mm]
                agb & \(\displaystyle v(r) = \vinit + \frac{\vinf - \vinit }{1 + \left ( \frac{\rmid}{R_*} \right )^{\alpha} \left ( \frac{r}{R_*} \right )^{-\alpha}} \) & AGB winds\\ [5mm]
                beta\_law & \(\displaystyle v(r) = \vinit + (\vinf - \vinit) \left ( 1 - \frac{R_*}{r} \right )^{\beta} \) & Hot/cool  \\
                & & stars\\
                \end{tabular}
                \end{adjustbox}
            \end{table}

        % Accelerate following a given velocity law using bridge
        Near the surface of the star, usually within a few stellar radii, the wind is accelerated to the terminal wind velocity.
        In the accelerating wind mode, the wind particles are created in the same way as in the simple wind mode, but with a lower velocity, $v < \vinf$.
        All particles near the star are artificially accelerated in such a way that the wind follows a predefined velocity profile.

        % Bridge works like... and we use it to ...
        The artificial acceleration is implemented using \bridge\ \citep{fujii_bridge:_2007}.
        Originally, \bridge\ was designed to couple multiple gravitational codes.
        In this method, each code is evolved separately for a short, predefined timestep\footnote{
            This should not be confused with the internal timesteps for each code which may be variable, meaning particles have different timesteps depending on conditions such as local particle density.}.
        The mutual gravitational effect is included by \bridge\ using a kick-drift-kick scheme \citep[see e.g.][]{AMUSE_book}.
        This method can also be used to gravitationally couple a pure N-body code with an SPH code, or to apply a gravitational potential to the particles in one or more codes.
        In \stellarwind, we use \bridge\ by including an artificial potential near the star, and then use the same kick-drift-kick scheme to ensure a smooth acceleration of the wind particles.

        % We have implemented ..., functions and more can be added
        In Figure~\ref{fig: acc functions} and Table~\ref{tab: acc functions} we present the acceleration functions (sometimes referred to as acceleration laws) currently implemented in \stellarwind.
        For a given velocity (or acceleration) profile, all required quantities are calculated following the equations in Appendix~\ref{sec: equations}.
        The constant velocity function is similar to the simple wind mode in that when the wind particles are created, they already have the terminal velocity.
        However, as noted below, when used in the accelerating mode we can add extra terms to compensate for the gravity of the star, as well as the pressure of the gas on the wind.
        These extra accelerating terms are added after the particles have been created, which is not possible with the simple wind mode.
        In this way, we guarantee the desired constant velocity profile.
        The logistic and agb functions provide a fit to the time-averaged behaviour of dynamical models of asymptotic giant branch (AGB) winds from \cite{nowotny_line_2010}.
        These winds exhibit a specific acceleration zone, the location of which can be chosen with the parameters \rmid\ and either s or $\alpha$ (for the logistic and agb function, respectively).
        These parameters determine the center and the width of the acceleration zone.
        The default values $\rmid = 3$ and  $s = \alpha = 10$ are chosen to fit the dynamic models.
        The beta\_law function, which was derived using a combination of observations and theoretical wind models, was taken from \citet{lamers_introduction_1999} and \citet{maciel_hydrodynamics_2014}.
        The $\beta$ parameter indicates the steepness of the acceleration curve and is often derived from observations.
        The example values $\beta = 0.8$ and $\beta = 2$ are typical for hot and cool stars respectively.
        The rsquared and delayed\_rsquared functions can be used as rough approximations to the wind profiles of hot stars and cool giants, respectively.
        They have the advantage of being computationally faster than the beta-law, agb and logistic functions.
        In the delayed\_rsquared model the parameter \raccstart\ (with a default value of 2) sets the lower boundary of the acceleration zone.
        The initial velocity \vinit, which is used in all functions except for constant\_velocity, is of the order of a few km/s due to microturbulence in the stellar atmosphere where the material is launched.
        We note that low values of \vinit\ result in high densities which lead to slow simulations, so in many cases a higher value of \vinit\ can be used as an approximation.
        In addition to these predefined functions, new user defined velocity functions can easily be incorporated.

        % correction for gravity, below sound speed, many compensation terms are needed (and available)
        When the gravity of the star is included, an additional acceleration term can be added to compensate for it and ensure that the wind particles follow the chosen velocity profile.
        The gas pressure can also exert an acceleration on the wind.
        We therefore provide the option to subtract the expected gas pressure acceleration (see Appendix~\ref{sec: gas pressure}) from the applied artificial acceleration.
        If we do not include a hydrodynamical simulation of the stellar interior, unphysical boundary effects near the surface of the star can influence the wind evolution or even prevent the wind from being launched.
        We therefore provide the option to create a ``staging shell'' near the star, generally at least twice the thickness of the newly created shells.
        Within this shell, the accelerations are chosen in such a way that the desired velocities are enforced regardless of the gas dynamics.
        This shell then provides a better boundary condition for the rest of the simulation.

        % vacuum initial conditions
        For many simulations using the simple and accelerating wind we can start the simulation with a vacuum around the star into which the wind particles are released.
        However, this can lead to an extra acceleration at the outer radius as the vacuum does not exert any pressure on the outermost particles.
        This can be problematic, especially for slow winds, where this spurious acceleration can significantly increase the velocities.
        We therefore include a function in \stellarwind\ that creates an initial set of SPH particles following the desired temperature, density and velocity profiles up to a given radius.
        This function uses the same initial grid distribution described in Section~\ref{sec: SPH}.
        Since the whole grid is created at once, without random rotations between different shells, this can introduce preferential directions.
        We advise that any scientific measurements are started after all these particles have left the area of interest.

        % we can speed up particle creation using the critical timestep, but this is an approximation
        When particles are created, we ensure that they follow the desired density profile by solving equation~\ref{eq: x of v} for each particle.
        For most acceleration functions, this equation has to be solved numerically, which can severely slow down the simulation (see Section~\ref{sec: fw}).
        We therefore include the option to define a critical timestep, \tcrit.
        When new wind particles are created, $\delta t$, which determines the thickness of the shell of new particles, is compared to \tcrit.
        If $\delta t <$ \tcrit\ it means that the new wind particles have not reached the accelerating region yet.
        For this reason, the acceleration function is approximated by the constant velocity function, for which equation~\ref{eq: x of v} is solved analytically.
        This approximation is only valid for acceleration functions for which the velocity near the star is close to constant, like the logistic function, not for acceleration functions with a large acceleration near the stellar surface, like the beta\_law function (see Figure~\ref{fig: acc functions}).

    \subsection{Heating wind}
        \label{sec: heating}
        % For large scale, mainly general heating of interest
        The third wind mode is based on the method used in \citet{pelupessy_evolution_2012} and is designed for use in large scale simulations, e.g. embedded star clusters.
        For these simulations, the main effect of the wind is that it adds mass and energy to the surrounding gas, therefore this mode cannot be used for a star in a vacuum.
        Studying the detailed kinematics of the wind near the star is not the goal of these simulations and therefore a simpler approximation of the wind interaction is used.
        The advantage of this approximate approach is that it can be used at far lower resolution (longer timesteps and higher SPH particle mass) which greatly speeds up the simulations.
        If particles were created with a high velocity, small timesteps would be required to completely sample the particle trajectory and interactions with other particles.

        % We integrate the mechanical luminosity
        The basic idea of the heating wind mode is that new wind particles do not have an initial velocity relative to the star.
        Instead they have an internal energy, $u$, which corresponds to the mechanical energy (\Emech) of the accumulated wind, defined as,
        \begin{equation}
        \begin{split}
            \Emech & = \int_{t_0}^{t_1} \Lmech(t) \, dt, \\
            \Lmech(t) & = \frac{1}{2} \Mdot(t) \vinf(t)^2,
            \end{split}
            \end{equation}
        where \Lmech\ is the instantaneous mechanical luminosity and $t_0$ and $t_1$ are the previous and current wind release time respectively.
        The integral is numerically approximated in \stellarwind\ during the simulation.
        The internal energy of the new particles is set to
        \begin{equation}
            u = f_{\mathrm{fb}} \frac{\Emech}{\Delta M_*},
        \end{equation}
        where $\Delta M_*$ is the mass lost and $f_{\mathrm{fb}}$ is the feedback efficiency parameter that accounts for radiative losses.
        Typical values for these parameters can be found in the examples shown in Sections \ref{sec: em} and \ref{sec: sn}.

        % in case of a supernova we just add all the energy
        As discussed in \citet{pelupessy_evolution_2012}, this method of creating particles with appropriate internal energy can also be used to simulate a supernova.
        If a star goes supernova, the calculated mechanical energy is ignored, and instead $10^{51}$~erg of energy is divided over the newly created particles.
        It should be noted that the injection of so much energy in the surrounding gas will cause the gas to evolve very rapidly, which can lead to time-stepping artefacts \citep[e.g.][]{pelupessy_evolution_2012}.
        One way to prevent this is to use a very small timestep (preliminary tests suggest $\unsim 10$~yr) shortly after a supernova.

\section{Application}
        \label{sec: Tests}
        \begin{figure}[!tbp]
            \includegraphics[width=0.5\textwidth]{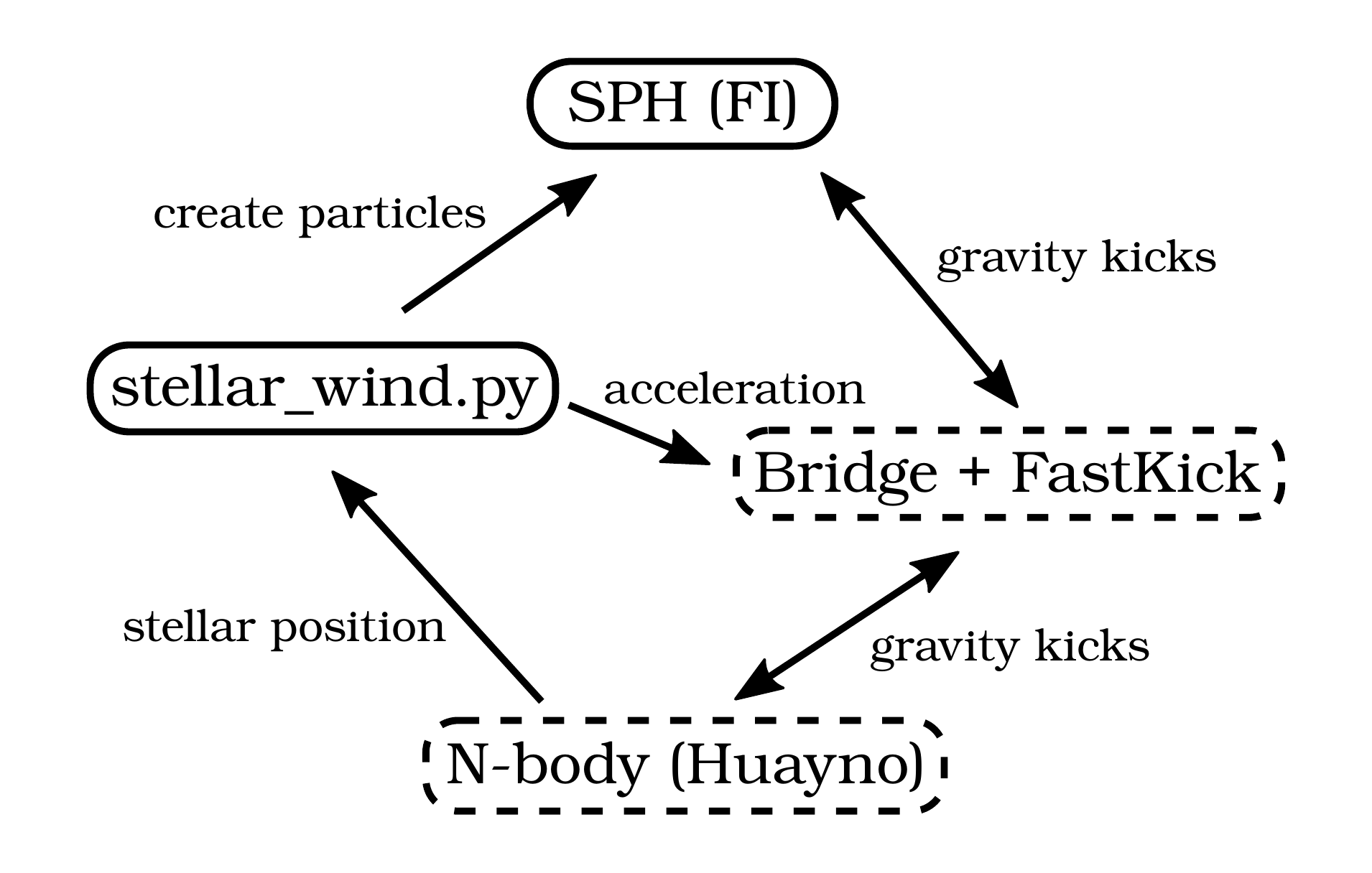}
            \caption{
                A diagram of the \amuse\ codes (boxes) and their relations (labelled arrows).
                Dotted lines indicate optional codes that can be added depending on the astrophysical application.
                \label{fig: flow diagram}}
            \end{figure}

        % We do ... tests with ... modes
        To ensure that \stellarwind\ performs as expected, we run test simulations with different initial conditions and wind modes.
        In this section we present the results of these tests.
        The tests in Sections~\ref{sec: fw}~and~\ref{sec: sw} are simulations of processes that happen close to the star.
        Therefore only the simple and accelerating wind modes are applicable.
        The tests in Section~\ref{sec: em}~and~\ref{sec: sn} are large scale simulations of the interaction between the stellar wind and the gas in which the star is embedded.
        For this type of simulation the heating wind mode is applicable.
        The final test in Section~\ref{sec: tr}, where we couple stellar dynamics, hydrodynamics and stellar winds, shows the power of \stellarwind\ within \amuse\ by simulating the colliding winds from a stable triple star system.
        The initial distribution of the wind particles is the one based on a grid for all the models in these tests.
        The particular parameters used are described accordingly in the following subsections.

        % For these simulations we use the codes ... with numerical parameters ...
        The \stellarwind\ module is designed to couple different parts of a simulation in \amuse.
        Testing the code requires the use of other simulation codes, which have their own parameters to be set.
        Here we describe the general method and general parameters used in our test models.
        In Figure~\ref{fig: flow diagram} we present a flow diagram to illustrate the codes and the relationships between them.
        Note that the codes and coupling strategies used here are merely an example and should be modified in order to be suitable for any specific application.
        To simulate the gas we use the SPH code \ficode\ with an adiabatic equation of state and artificial viscosity parameters $\alpha = 0.5$ and $\beta=1.0$ \citep[following][]{lombardi_tests_1999}.
        Self-gravity of the gas is only included in the tests which do not require periodic boundary conditions, i.e. those tests where the simple or accelerating mode are used.
        For simulating the gravitational attraction of the star on the gas we need \bridge, which is also used in the accelerating wind mode (Section \ref{sec: accelerating}).
        The \bridge\ code requires an additional code for calculating the gravitational force.
        When we simulate only a single star that does not evolve dynamically, we use \fastkick
            \footnote{The \fastkick\ code, developed by N. de Vries, is an unpublished gpu-enabled code in \amuse\ that can calculate the gravitational force of one set of particles on another set of particles.
                      It is ideal for the gravitational coupling between particles in different codes via \bridge.}.
        When we simulate multiple stars that evolve dynamically (Section~\ref{sec: tr}), we use the N-body code \huayno\ \citep{pelupessy_n-body_2012}.
        For each simulation, we also need to define a number of integration timescales, such as the \bridge\ timestep ($t_{\mathrm{br}}$), the (maximum) internal timestep (\tnbody) of the SPH and N-body codes and the wind release timestep ($t_\mathrm{wind}$), as well as the end time ($t_\mathrm{end}$) of the simulation.
        The choice of these timesteps depends on the problem and the type of simulation.
        For the bridge leap-frog algorithm to work, we should set $t_{\mathrm{br}} \ge \tnbody$ and the wind code requires $t_{\mathrm{wind}} \ge t_{\mathrm{br}}$.
        It is also a good idea to ensure that larger timescales are integer multiples of smaller timescales.
        For the simulations in this paper, we only define $t_{\mathrm{wind}}$ and choose $t_{\mathrm{wind}} = 2 t_{\mathrm{br}} = 4 \tnbody$.
        For the hydro simulation we also need to set the SPH particle mass ($M_\mathrm{SPH}$).

    \subsection{Fast winds}
        \label{sec: fw}

        \begin{table}
            \caption{
                Stellar and wind parameters used in the fast wind test.
                Derived parameters are indicated with an arrow ($\rightarrow$).
                Since the smoothing length is highly variable with extreme outliers, we include the median value of all gas particles shown in Figure~\ref{fig: fw vel dens}.
                \label{tab: fw parameters}}

            \begin{tabular}{l l l}
                \B name & parameter & value \\
                \hline
                \T mass-loss rate & \Mdot & $10^{-6}$~\MSun/yr \\
                terminal wind velocity & \vinf & 700~km/s \\
                initial wind velocity & \vinit & 100~km/s \\ [2mm]
                stellar mass & $M_*$ & 20~\MSun \\
                stellar radius & $R_*$ & 30~\RSun \\
                stellar luminosity & $L_*$ & 100\,000~\LSun \\
                stellar surface temperature & $T_*$ & 20\,000~K \\
                escape velocity at $R_*$ & $v_\mathrm{esc}(R_*)$ & $\rightarrow 504.5$~km/s\\  [2mm]
                wind timestep & $t_\mathrm{wind}$ & 0.02~days \\
                end time & $t_\mathrm{end}$ & 5~days \\
                SPH particle mass & $M_\mathrm{SPH}$ & $10^{-11}$~\MSun \\
                particles per shell & $N_\mathrm{shell}$ & $\rightarrow \unsim 5$ \\
                particles in simulation & $N_\mathrm{tot}$ & $\rightarrow \unsim 1378$ \\
                median smoothing length & $h$ & $\rightarrow \unsim 32$~\RSun \\
                \end{tabular}
            \end{table}

        \begin{figure}[!tbp]
            \includegraphics[width=0.5\textwidth]{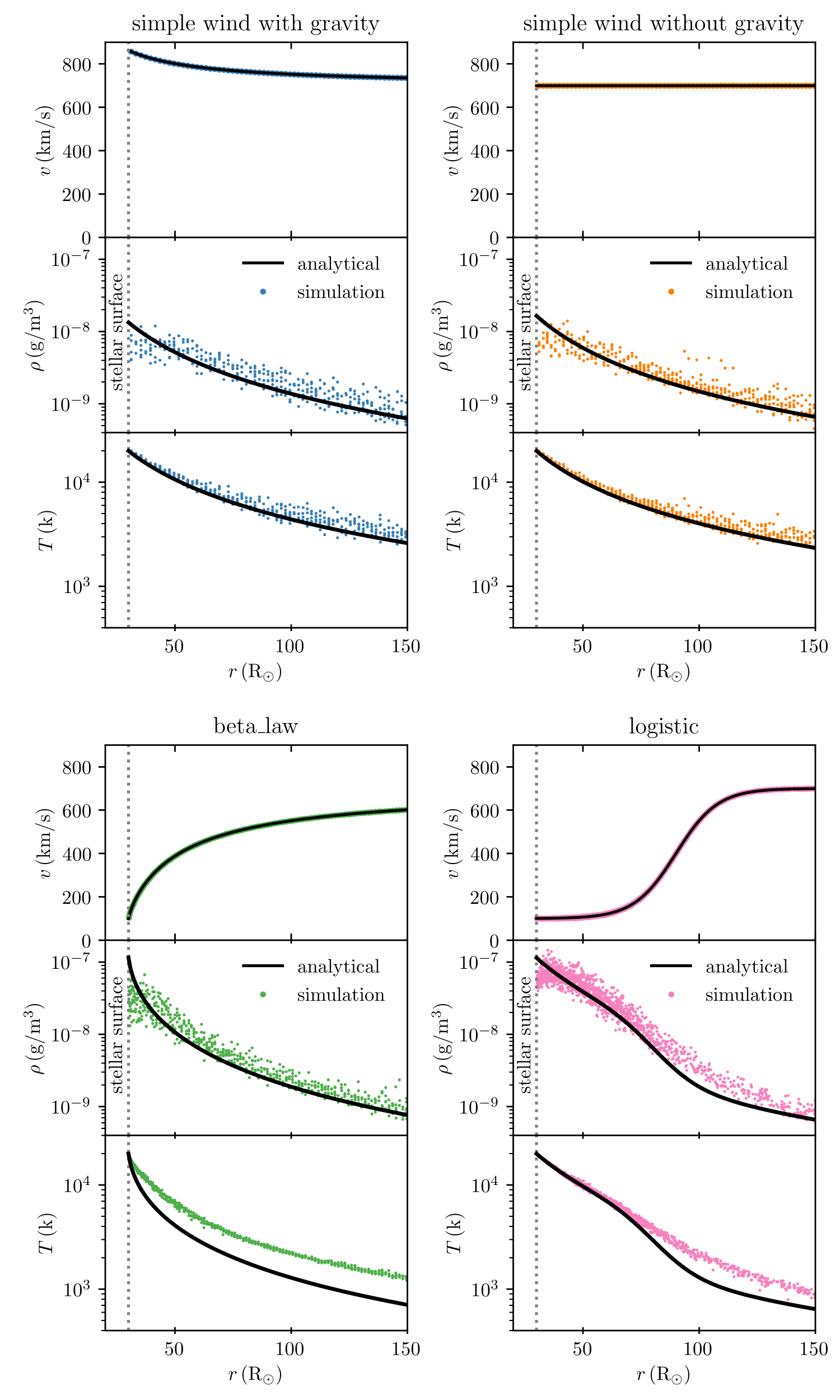}
            \caption{
                The analytical and simulated velocity ($v$), density ($\rho$) and temperature ($T$) as a function of radius ($r$) for the fast wind test.
                The results are from simulations using the simple wind mode (top) and the accelerating wind mode (bottom).
                The stellar and wind parameters can be found it Table~\ref{tab: fw parameters}.
                To calculate the analytical temperature profile, we assume adiabatic expansion.
                \label{fig: fw vel dens}}
            \end{figure}

        \begin{figure}[!tbp]
            \includegraphics[width=0.5\textwidth]{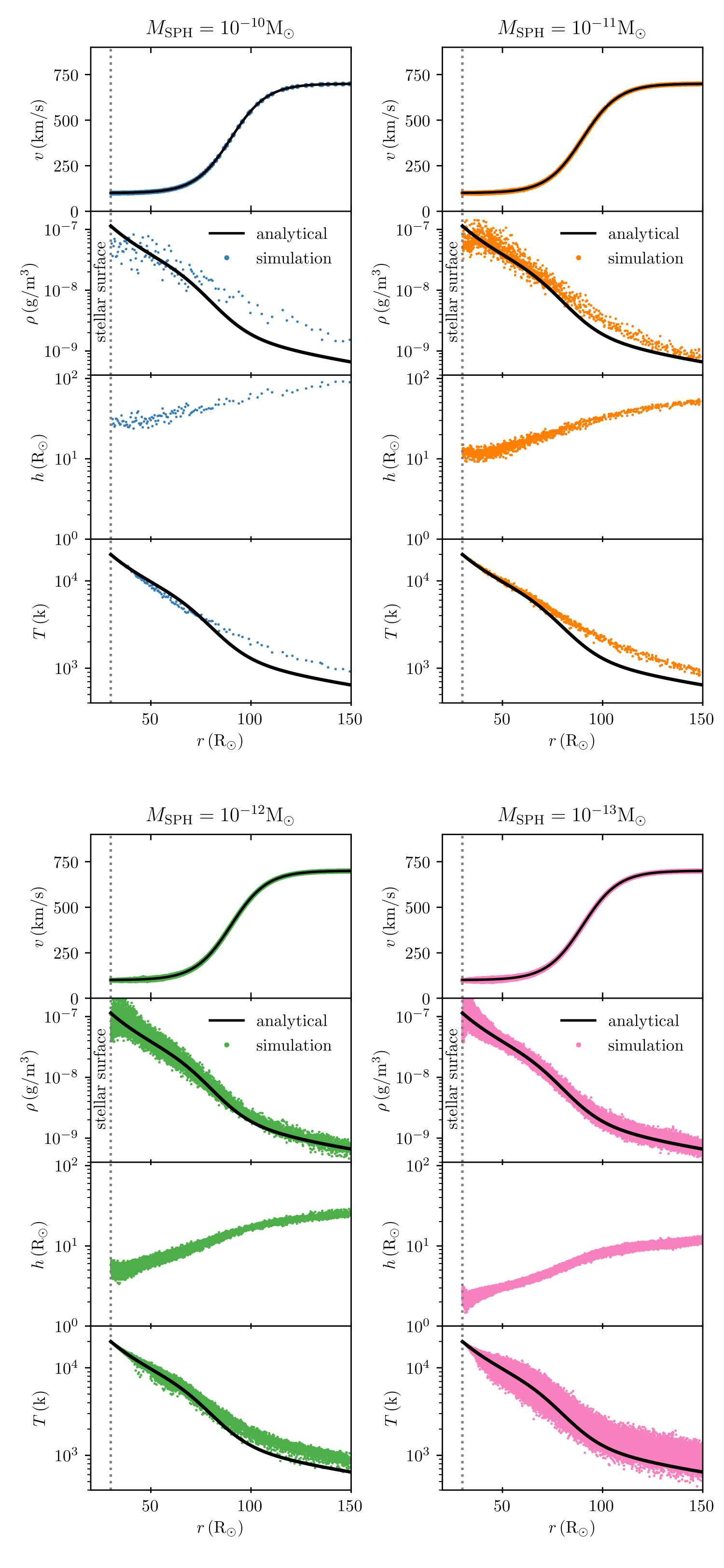}
            \caption{
                The same as Figure~\ref{fig: fw vel dens}, but only for a simulation using the accelerating wind mode with the logistic acceleration function.
                We have varied the resolution by changing the SPH particle mass ($M_\mathrm{SPH}$) and through that the number of particles in the simulation.
                We have added the smoothing length, $h$, as a function of radius for each simulation, which is a measure of the local spatial resolution.
                \label{fig: fw convergence}}
            \end{figure}

        \begin{figure}[!tbp]
            \includegraphics[width=0.5\textwidth]{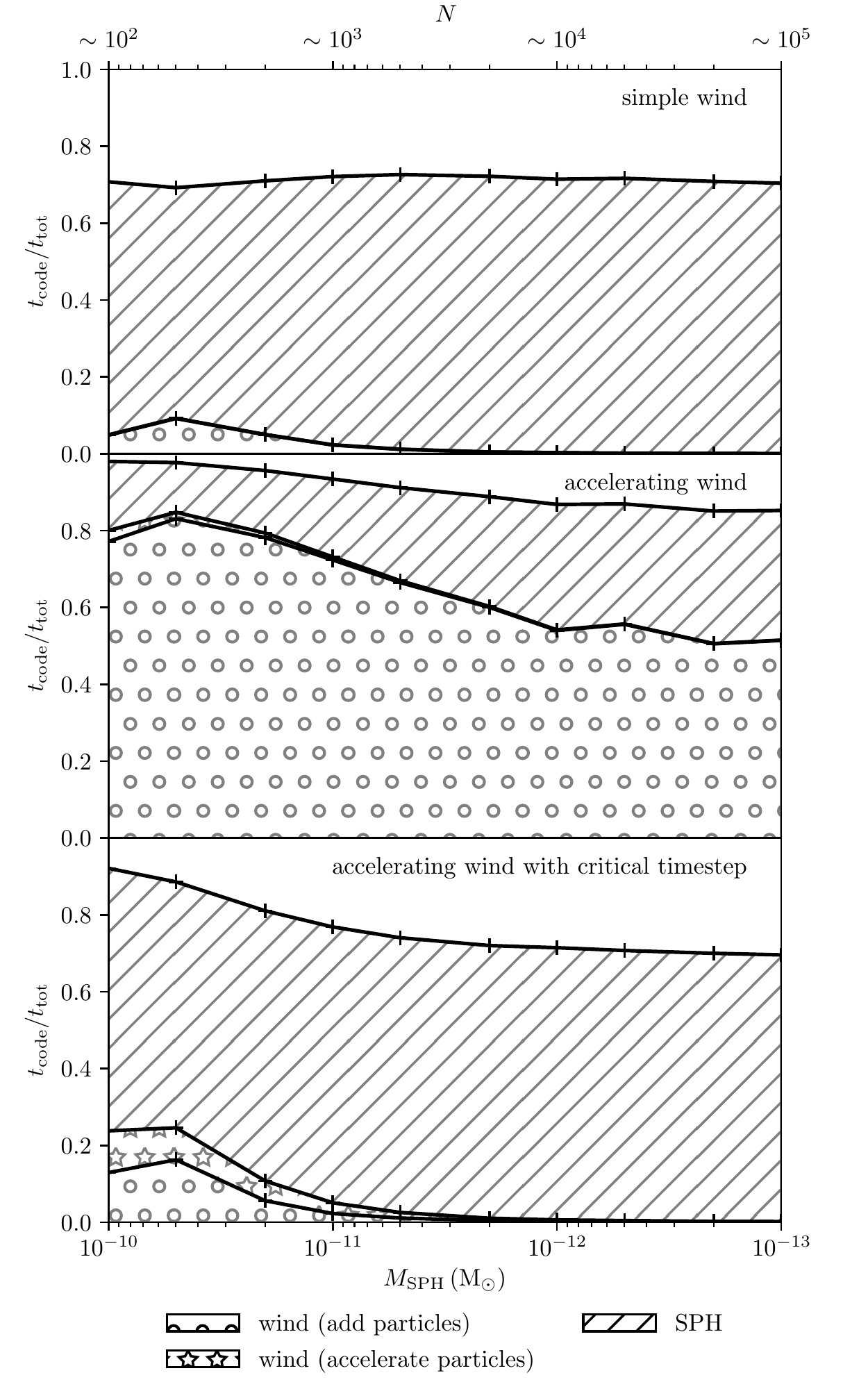}
            \caption{
                The part of the simulation time used for the \stellarwind\ code while creating new particles (circles) and accelerating them (stars) compared to the time used by the SPH code (diagonal lines).
                The three panels show results for simulations with the simple wind mode (top), the accelerating wind mode with the logistic acceleration function without a critical timestep (middle) and the accelerating wind mode with a critical time step (bottom).
                Marks along each line denote separate simulation runs.
                At the top axis we give an estimate of the number of SPH particles ($N$) actually used in the simulation with the corresponding particle mass ($M\sph$).
                The remaining simulation time (white space) was mostly spent on unoptimized administrative tasks like saving snapshots and removing escaping particles.
                \label{fig: fw scaling}}
            \end{figure}

        % The setup is similar to a hot massive luminous star (with table) the wind is supersonic
        In this section we present the results of a set of simulations using the simple and accelerating wind modes.
        We simulate the wind from a single, hot, massive, luminous star, for which we present the parameters in Table~\ref{tab: fw parameters}.
        Note that these values were not chosen using a specific stellar model and this test should only be considered as an example of the use of \stellarwind.
        The initial wind velocity, $\vinit = 100$~km/s, is based on numerical considerations.
        As we shall see in Section~\ref{sec: sw}, slow (subsonic\footnote{If the wind speed near the star is lower than the local sound speed, the wind is called subsonic. On the other hand if the wind speed is higher than the local sound speed, i.e. past the 'sonic point', the wind is called supersonic.}) winds are more complex to simulate and for many applications using a higher \vinit\ is sufficient.

        % In figure ..., we see that the wind follows the desired profiles well (including convergence test).

        In Figure~\ref{fig: fw vel dens} we show the velocity, density and temperature profiles for the simple wind mode and two accelerating functions: the beta\_law and logistic function.
        In all cases the velocity profiles follow the desired analytical velocity curve.
        For the simple wind mode, the density and temperature profiles also follow the desired curve, but with more scatter.
        For the accelerating wind curves, we see that in regions with high acceleration the densities and temperatures in the simulation are too high.
        This is a result of the low resolution in combination with the way densities are calculated in SPH, using a kernel function that 'smears out' these variables.
        We show in Figure~\ref{fig: fw convergence} that for a higher resolution (smaller $M\gas$), the desired density and temperature curve are recovered.
        Note that the logistic acceleration function is not a good representation for the velocity profile of a hot massive luminous star.
        This example was chosen to illustrate the discrepancies that can potentially occur.
        For any scientific application of this code, a detailed convergence test for the selected setup will still be required.

        % In figure ..., we see how runtime percentage that the wind code takes, depends on N
        In addition to being accurate, it is also important that a simulation code is fast.
        In Figure~\ref{fig: fw scaling} we present the time spent in different parts of the simulation code (\tcode) divided by the total cpu (or wall-clock) time (\ttot) as a function of resolution\footnote{These simulations where all performed on the same desktop computer using a 4-core Intel Xeon E5507 CPU.}.
        In the top panel we see that when using the simple wind mode, the time spent in the \stellarwind\ code is less than 1\% when using more than $\unsim 10^4$ particles.
        Most of the simulation time is therefore spent in the SPH code itself, which is what we want.
        When we use the accelerating wind mode however (middle panel), the particle creation becomes a major bottleneck because numerically solving equation \ref{eq: x of v} is slow.
        To speed up the simulation, we have included the option to approximate the acceleration function with a constant velocity when particles are created near the star by defining a critical timestep (\tcrit, see Section~\ref{sec: accelerating}).
        In the bottom panel we see that by using this approximation, the time spent in \stellarwind\ reduces to $< 1$\% for $> 10^4$ particles.

    \subsection{Slow wind}
        \label{sec: sw}

        \begin{table}
            \caption{
                Stellar and wind parameters used in the slow wind test.
                Derived parameters are indicated with an arrow ($\rightarrow$).
                Since the smoothing length is highly variable with extreme outliers, we include the median value of all gas particles shown in Figure~\ref{fig: sw acc}.
                \label{tab: sw parameters}}

            \begin{tabular}{l l l}
                \B name & parameter & value \\
                \hline
                \T mass-loss rate & \Mdot & $5 \cdot 10^{-7}$~\MSun/yr \\
                terminal wind velocity & \vinf & 25~km/s \\
                initial wind velocity & \vinit & 2~km/s \\ [2mm]
                stellar mass & $M_*$ & 2~\MSun \\
                stellar radius & $R_*$ & 300~\RSun \\
                stellar luminosity & $L_*$ & 8\,000~\LSun \\
                stellar surface temperature & $T_*$ & 3\,000~K \\
                escape velocity at $R_*$ & \vesc($R_*$) & $\rightarrow$ 50.45 km/s \\[2mm]
                wind timestep & $t_\mathrm{wind}$ & 2~days \\
                end time & $t_\mathrm{end}$ & 2\,000~days \\
                SPH particle mass & $M_\mathrm{SPH}$ & $10^{-9}$~\MSun \\
                particles per shell & $N_\mathrm{shell}$ & $\rightarrow \unsim 5$ \\
                particles in simulation & $N_\mathrm{tot}$ & $\rightarrow \unsim 8476$ \\
                median smoothing length & $h$ & $\rightarrow \unsim 159$~\RSun \\
                \end{tabular}
            \end{table}

        \begin{figure}[!tbp]
            \includegraphics[width=0.5\textwidth]{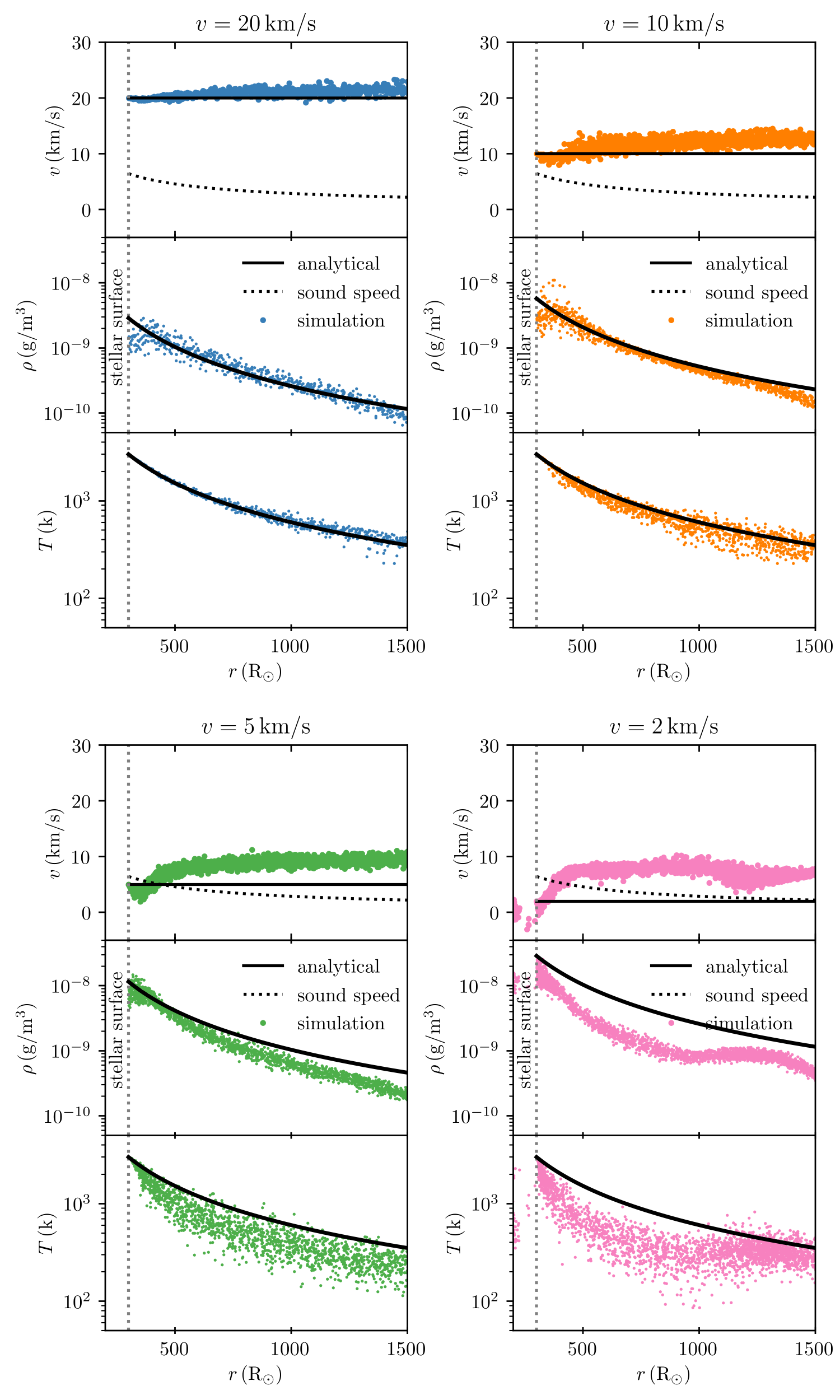}
            \caption{
                Same as Figure~\ref{fig: fw vel dens} but for the slow wind simulations using the simple wind mode without gravity where we vary the wind velocity.
                We have added the expected local sound speed (dotted line) for comparison.
                \label{fig: sw simple}}
            \end{figure}

        \begin{figure}[!tbp]
            \includegraphics[width=0.5\textwidth]{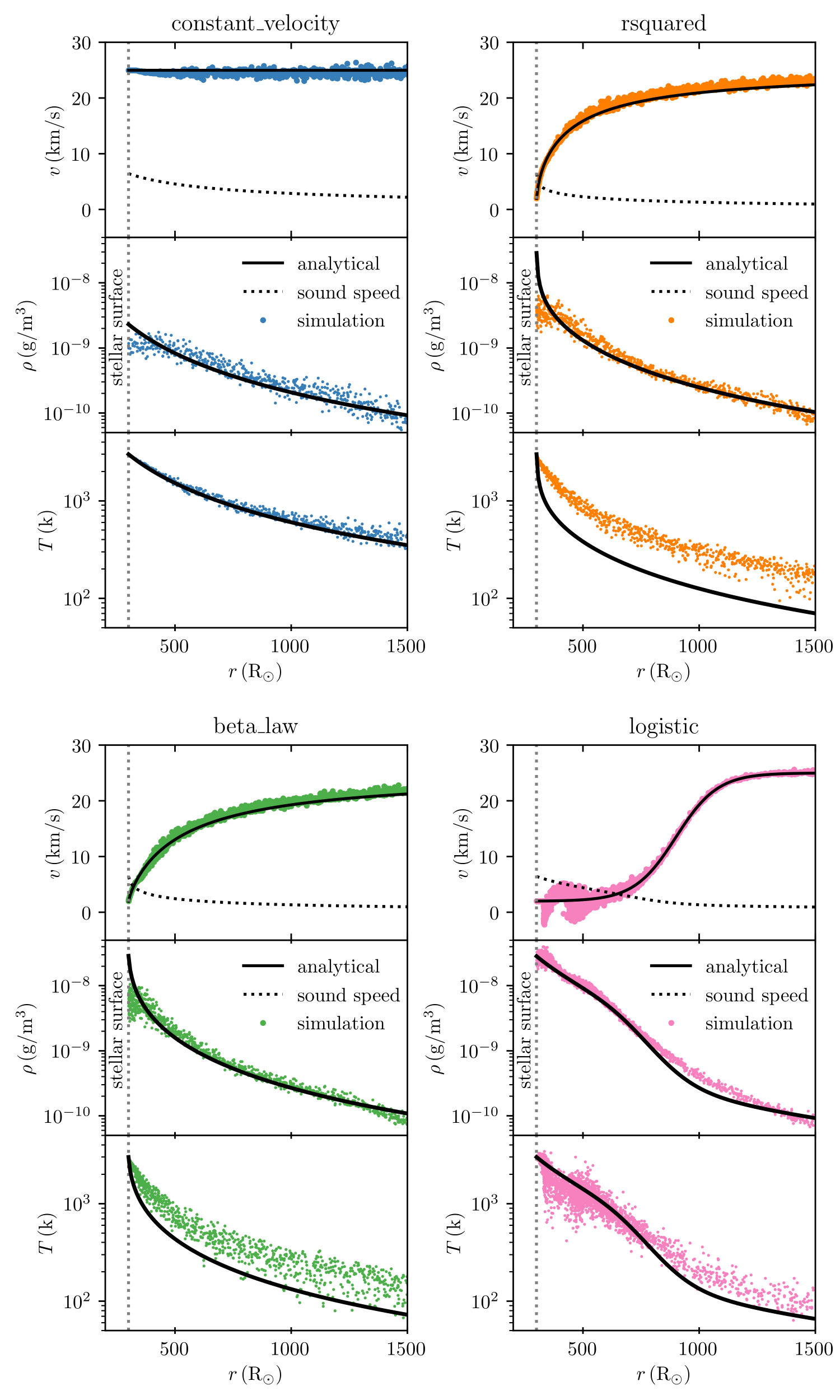}
            \caption{
                Same as Figure~\ref{fig: sw simple} but for the accelerating wind mode with four different acceleration functions.
                \label{fig: sw acc}}
            \end{figure}

        \begin{figure}[!tbp]
            \includegraphics[width=0.5\textwidth]{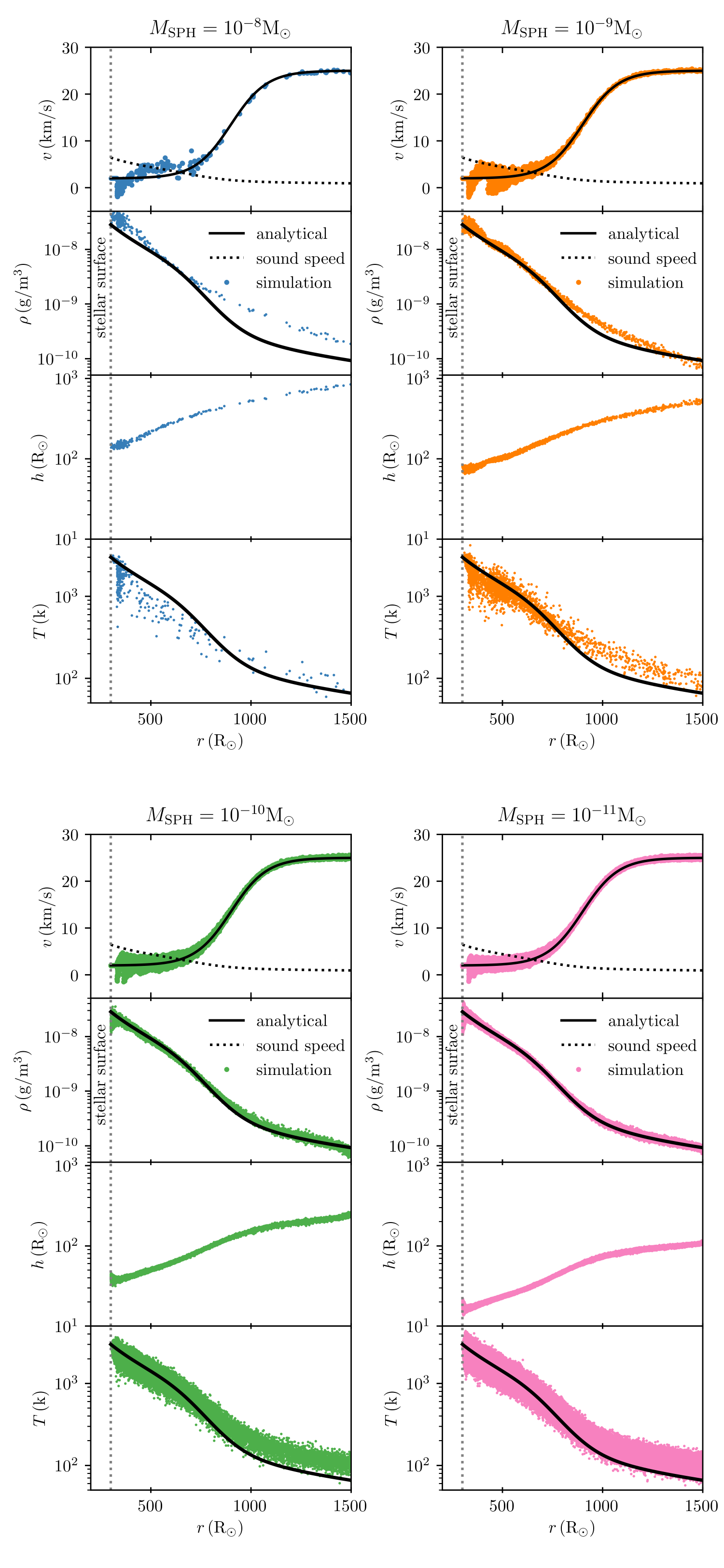}
            \caption{
                Same as Figure~\ref{fig: fw convergence} but for the slow wind test.
                \label{fig: sw convergence}}
            \end{figure}

        % Table: initial parameters

        % If wind is subsonic, hydrodynamics comes and makes it more difficult
        For the slow wind test, we simulate the wind from a single, cool, giant star, for which we present the parameters in Table~\ref{tab: sw parameters}.
        The values of these parameters are not computed with a stellar evolution code, but they correspond to typical values for AGB stars.
        Part of the stellar wind is subsonic and therefore hydrodynamical effects are no longer negligible, unlike in the fast wind test.

        % In figure ..., we see that simple wind is not good enough
        In Figure~\ref{fig: sw simple} we present the velocity profiles of simulations using the simple wind mode and varying \vinf\ at $t = 5$~days.
        We have not included the stellar gravity in these simulations, therefore the escape speed is not a relevant factor.
        However, when the wind speed is near or below the sound speed, the gas pressure gradient dominates and affects the terminal wind velocity.
        For very low wind speeds, this can cause wind particles to move inside the star, which is unphysical.
        We therefore conclude that the simple wind mode is not reliable for slow winds.

        % for figure ..., we  did
        In Figure~\ref{fig: sw acc} we present the results of simulations using the accelerating wind mode.
        In these simulations we have used the staging shell option (Section~\ref{sec: accelerating}) that enforces the correct particle velocity for all SPH particles with radius $r < 1.1 R_*$.
        We have also used the option to subtract the expected gas pressure acceleration from the acceleration to ensure that the particles follow the desired velocity profile.
        To avoid spurious acceleration of the outer particles, we start the simulation after initially creating a sphere of particles following the desired velocity and density profiles throughout the simulation area (up to $r = 1500 \RSun$).

        % we see that with all corrections, the accelerating wind does work.
        For the constant, rsquared and beta\_law velocity profiles, the particles follow the desired velocity profiles.
        For the logistic velocity profile, where particles are subsonic for a longer time, the simulated velocity profile deviates from the desired velocity profile in the subsonic region.
        However, the corrections described above ensure that the resulting velocities after particles pass the sonic point follow the desired velocity profile.
        To see how this deviation will affect the results of simulations where the subsonic region is of interest, we have compared this velocity profile with detailed velocity profiles of AGB stars \citep{nowotny_line_2010}.
        In these profiles, the velocities in the subsonic region oscillate due to stellar pulsations and do not follow the simple acceleration functions we have used here.
        We therefore advise caution when interpreting results of simulations in the subsonic region.

        % Density profile is again a resolution thing.
        Similar discrepancies in density and temperature as seen for the fast wind test in Figure~\ref{fig: fw vel dens} are also present for the slow wind test in Figure~\ref{fig: sw acc}.
        In Figure~\ref{fig: sw convergence} we present the results of a resolution test for the slow wind test.
        We see that the discrepancies in density and temperature decrease with higher resolution, as expected.
        The deviations in the velocity profile in the subsonic region also decrease, although some differences are still present even at high resolution.

    \subsection{Embedded star}
        \label{sec: em}

        \begin{table}
            \caption{
                The parameters used in the embedded star test.
                The stellar and wind parameters are the same as in Table~\ref{tab: fw parameters}.
                \label{tab: em parameters}}

            \begin{tabular}{l l l}
                \B name & parameter & value \\
                \hline
                \T gas density & $\rho\gas$ & 10 and 100~\MSun/pc$^3$ \\
                wind timestep & $t_\mathrm{wind}$ & $2 \cdot 10^4$~yr \\
                end time & $t_\mathrm{end}$ & $10^6$~yr \\
                SPH particle mass & $M_\mathrm{SPH}$ & 0.05 to 1~\MSun \\
                wind release radius & $r_\mathrm{wind}$ & 0.01~pc \\
                feedback efficiency & $f_{\mathrm{fb}}$ & 0.01 \\
                particles per shell & $N_\mathrm{shell}$ & $\rightarrow$ 0 or 1 \\
                new particles in simulation & $N_\mathrm{new}$ & $\rightarrow$ 5 to 100 \\
                median smoothing length & $h$ & $\rightarrow \unsim 0.2$~pc \\
                \end{tabular}
            \end{table}

        \begin{figure}[!tbp]
            \includegraphics[width=0.5\textwidth]{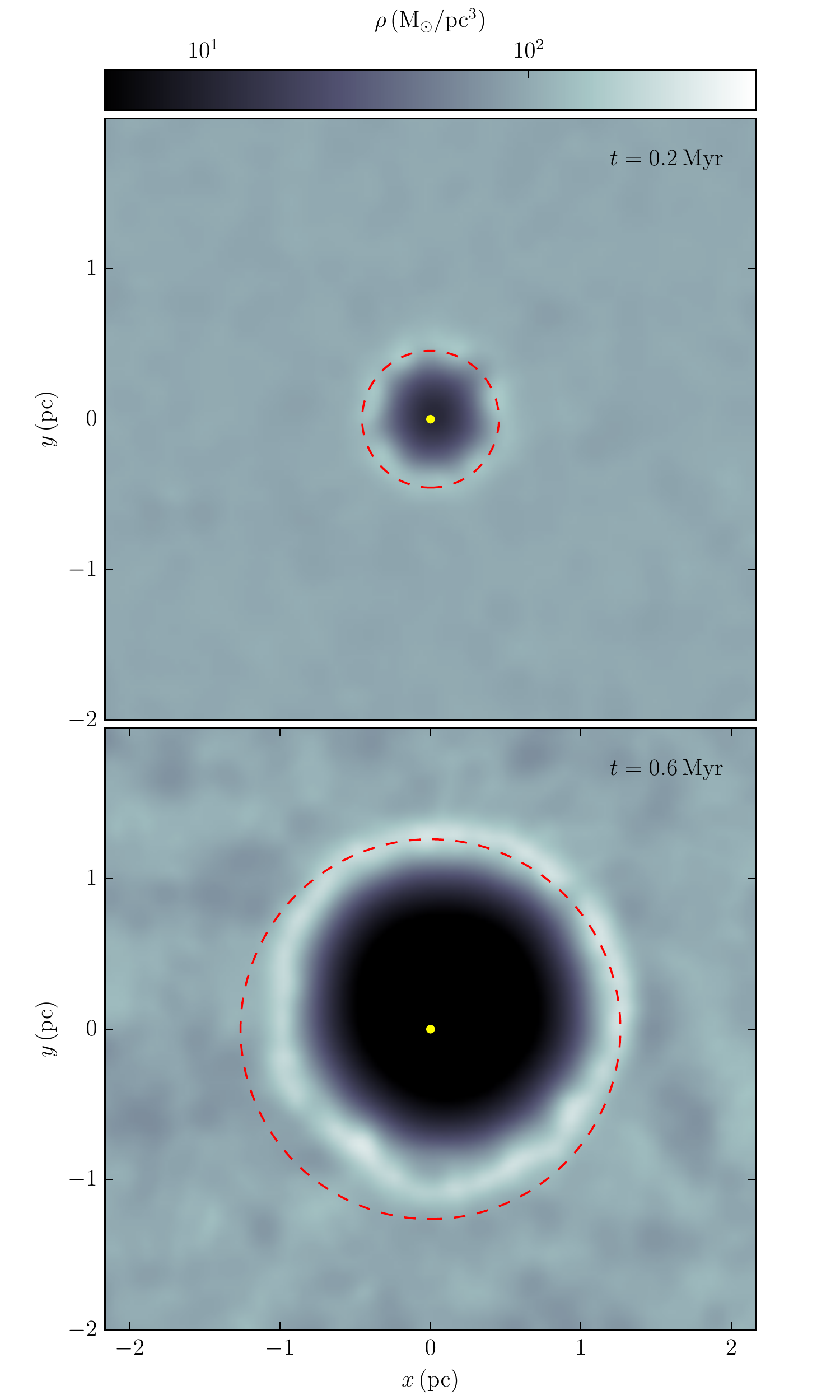}
            \caption{
                The gas density in the $x - y$ plane after 0.2~Myr (top) and after 0.6~Myr (bottom) for the embedded star simulation with $M_\mathrm{SPH}=0.1$~\MSun\ and $\rho\gas = 100$~\MSun/pc$^3$.
                The embedded star is positioned at the origin (yellow dot) and the red dashed circle shows the radius with the largest mean density.
                \label{fig: em density}}
            \end{figure}

        \begin{figure}[!tbp]
            \includegraphics[width=0.5\textwidth]{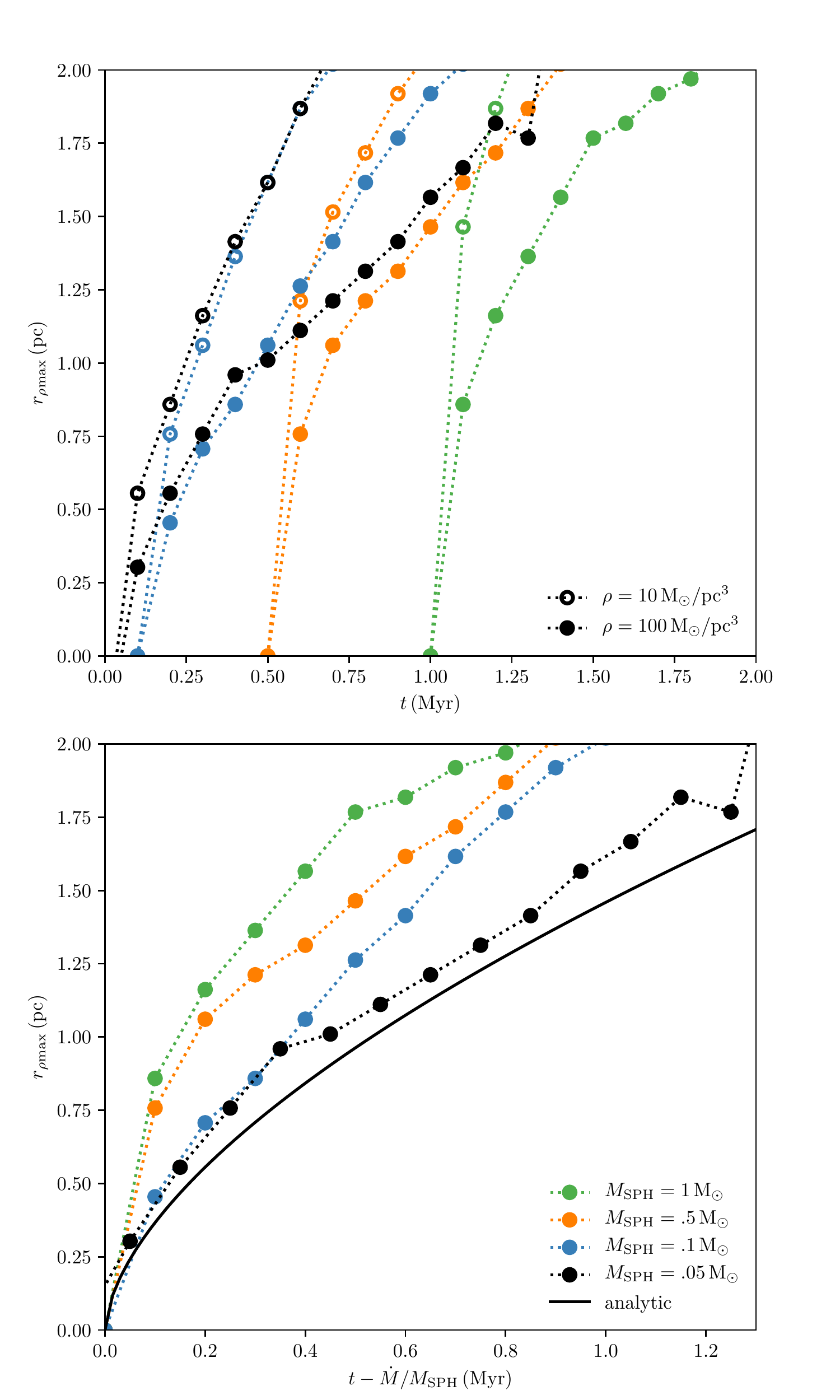}
            \caption{
                The radius with the highest mean gas density ($r_{\rho \mathrm{max}}$) as a function of time, $t$ for the embedded star simulations.
                Different colours correspond to different resolutions resulting from different SPH particle masses (M\sph).
                In the top panel, we show $r_{\rho \mathrm{max}}$ for all simulations as a function of $t$.
                Lines with open circles correspond to simulations where the gas density, $\rho\gas = 10$~\MSun/pc$^3$ and lines with filled circles to simulations with a gas density, $\rho\gas = 100$~\MSun/pc$^3$.
                In the bottom panel, we only show simulations with $\rho\gas = 100$~\MSun/pc$^3$ and subtract $\dot{M}/M_\mathrm{SPH}$ (the time of the first SPH particle creation) from $t$.
                The black solid line shows the analytical solution for the shell radius of an energy driven flow in a constant density medium \citep{dyson_interpretation_1984}.
                \label{fig: em dens radii}}
            \end{figure}

        % fw star at large scale, embedded in medium
        For the embedded star test (Table~\ref{tab: em parameters}), we take the hot, massive, luminous star from Section~\ref{sec: fw} and embed it in a constant density medium.
        The stellar wind will heat the gas and create a cavity around the star.
        This situation is quite common in embedded star clusters and it is what the heating wind mode is designed for.
        The initial gas is distributed along a grid\footnote{As mentioned in Section~\ref{sec: SPH}, this can introduce preferential directions and a glass or other relaxed system should be considered for most applications.} to ensure a constant density and a divergence-free random Gaussian velocity field following \citet{bonnell_hierarchical_2003}.
        To ensure that the medium is stable, we use periodic boundary conditions and stop the simulation when the wind-blown bubble covers more than half the simulation box.
        For the heating wind mode, the outer radius for new wind particles ($r_\mathrm{wind}$) is set manually to $r_\mathrm{wind} = 0.01$~pc and the feedback efficiency is set to $f_{\mathrm{fb}} = 0.01$ following \citet{pelupessy_evolution_2012}.

        % We see that a bubble is blown out
        In Figure~\ref{fig: em density} we show the gas density when the bubble has just started to form ($t=0.2$~Myr) and when it has had time to grow ($t=0.6$~Myr).
        The stellar wind creates an approximately spherical bubble of lower density as the gas is swept up in a high density shell around it.
        To understand why the bubble is not perfectly spherical, we note that the finite number of gas particles cause small numerical fluctuations in the initial gas density.
        When we then introduce a small number of wind particles with higher energy than the surrounding gas, these small fluctuations grow into a larger asymmetry in the wind bubble.
        This growth of small initial asymmetries was observed in SPH simulations of supernovae explosions \citep{rimoldi_simulations_2016} where they found that if the injected energy is spread out over more particles, the asymmetric effects diminish.
        If the asymmetry in the wind bubbles would become a problem for specific simulations, then the wind energy could be spread out in a similar way.

        % The convergence test shows that it even works for very high particle mass
        In Figure~\ref{fig: em density} we have drawn a dashed line that shows the radius where the mean density is highest ($r_{\rho \mathrm{max}}$, see Appendix~\ref{sec: dens radius} for details).
        At the start of the simulation, this radius is undefined, because the gas has a constant density.
        As the bubble grows and gas is swept up in an approximately spherical shell, the radius of maximum density matches the shell radius, which is what we plot as a function of time in Figure~\ref{fig: em dens radii}.
        Note that $r_{\rho \mathrm{max}}$ is slightly larger than the shell radius because of the asymmetry of the wind bubble.
        We present this expansion for different values of $M_\mathrm{SPH}$ (different resolutions) and two different gas densities.
        Even when the resolution is very low ($M_\mathrm{SPH} = 1.0~\MSun$), the heating wind method still results in a dispersion of the gas cloud.
        The expansion is faster for lower gas density, which is in agreement with analytical solutions for the shell radius of an energy driven flow in a constant density medium \citep{dyson_interpretation_1984}.
        However, the bubble expansion starts later for simulations with a lower resolution.
        This delay corresponds to the time it takes for the star to lose enough mass to create the first wind particle.
        For example, if $M_\mathrm{SPH} = 1.0~\MSun$ and $\Mdot = 1$~\MSun/Myr, this delay is 1~Myr.
        In the bottom panel of Figure~\ref{fig: em dens radii}, we show the shell expansion starting at the moment of the first wind injection.
        We see that lower resolution results in a faster expansion, caused by the larger energy injected in a single SPH particle.
        The expansion profile approaches the analytical solution for high resolution (small $M_\mathrm{SPH}$).
        We therefore advise that the choice of $M_\mathrm{SPH}$ be based on the stellar mass-loss rate and the delay and expansion profile that would be acceptable in the desired simulations.

    \subsection{Supernova}
        \label{sec: sn}

        \begin{table}
            \caption{
                The parameters used in the supernova test.
                \label{tab: sn parameters}}

            \begin{tabular}{l l l}
                \B name & parameter & value \\
                \hline
                \T
                initial stellar mass & $M_{\mathrm{ZAMS}}$ & 20~\MSun \\
                stellar age & $T_*$ & 9.78~Myr \\ [2mm]
                gas density & $\rho\gas$ & 10~\MSun/pc$^3$ \\
                end time & $t_\mathrm{end}$ & 2000~yr \\
                SPH particle mass & $M_\mathrm{SPH}$ & 0.01 \myendash 1~\MSun \\
                wind timestep & $t_\mathrm{wind}$ & 20~yr \\
                wind release radius & $r_\mathrm{wind}$ & 0.01~pc \\
                supernova energy & $E_{\mathrm{SN}}$ & $10^{51}$~erg \\
                mass-loss & $\Delta M_*$ & $12.95$~\MSun \\
                feedback efficiency & $f_{\mathrm{fb}}$ & 0.01 \\
                \end{tabular}
            \end{table}

        \begin{figure}
            \includegraphics[width=0.5\textwidth]{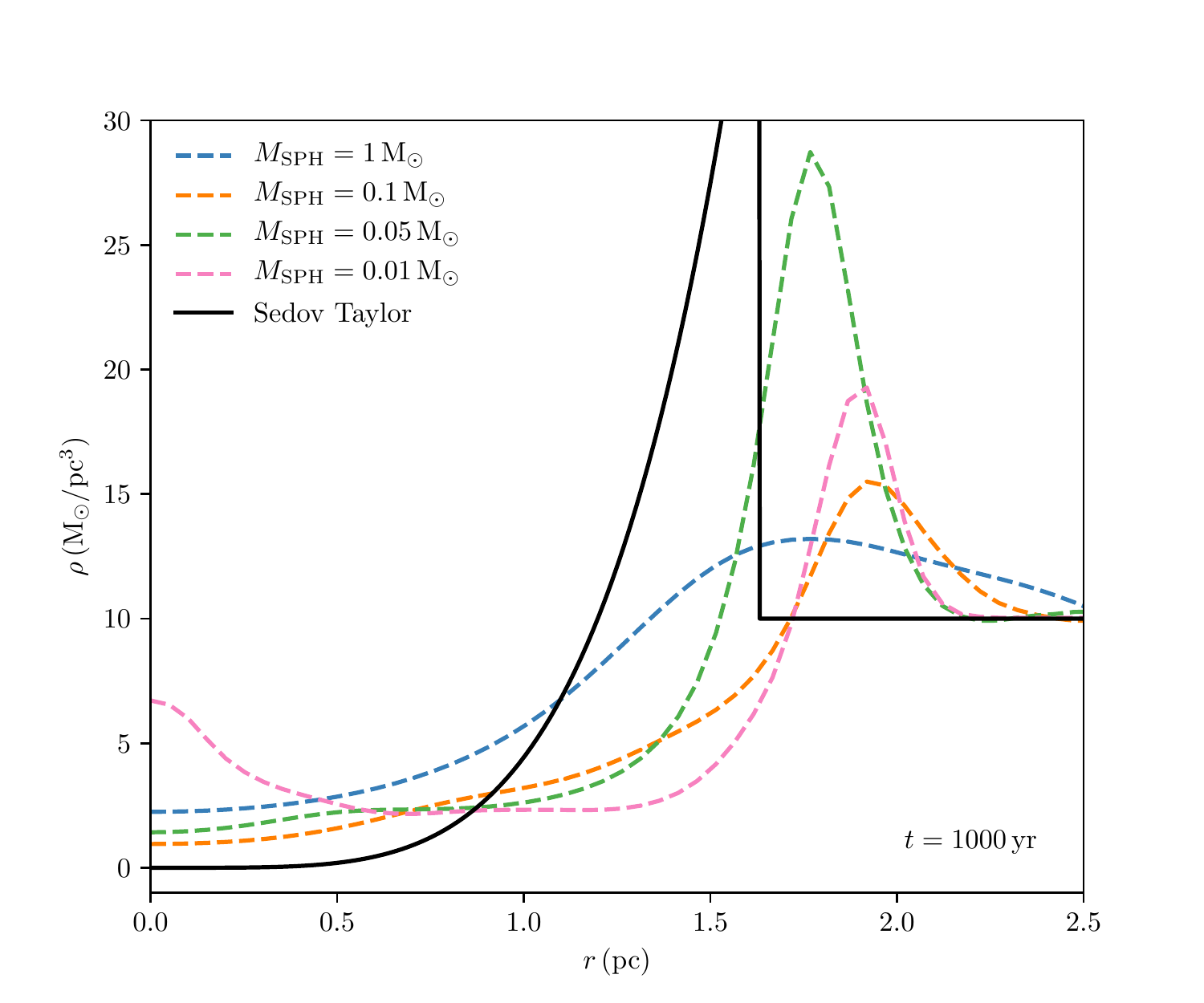}
            \caption{
                The mean gas density as a function of radius at time $t = 1000$~yr for the supernova test at four resolutions (dashed lines).
                The analytical solution (solid black line), is the Sedov-Taylor solution for a self-similar blast wave in a uniform medium \citep{taylor_formation_1950,sedov_similarity_1959}.
                \label{fig: sn dens profile}}
            \end{figure}

        \begin{figure}[!tbp]
            \includegraphics[width=0.5\textwidth]{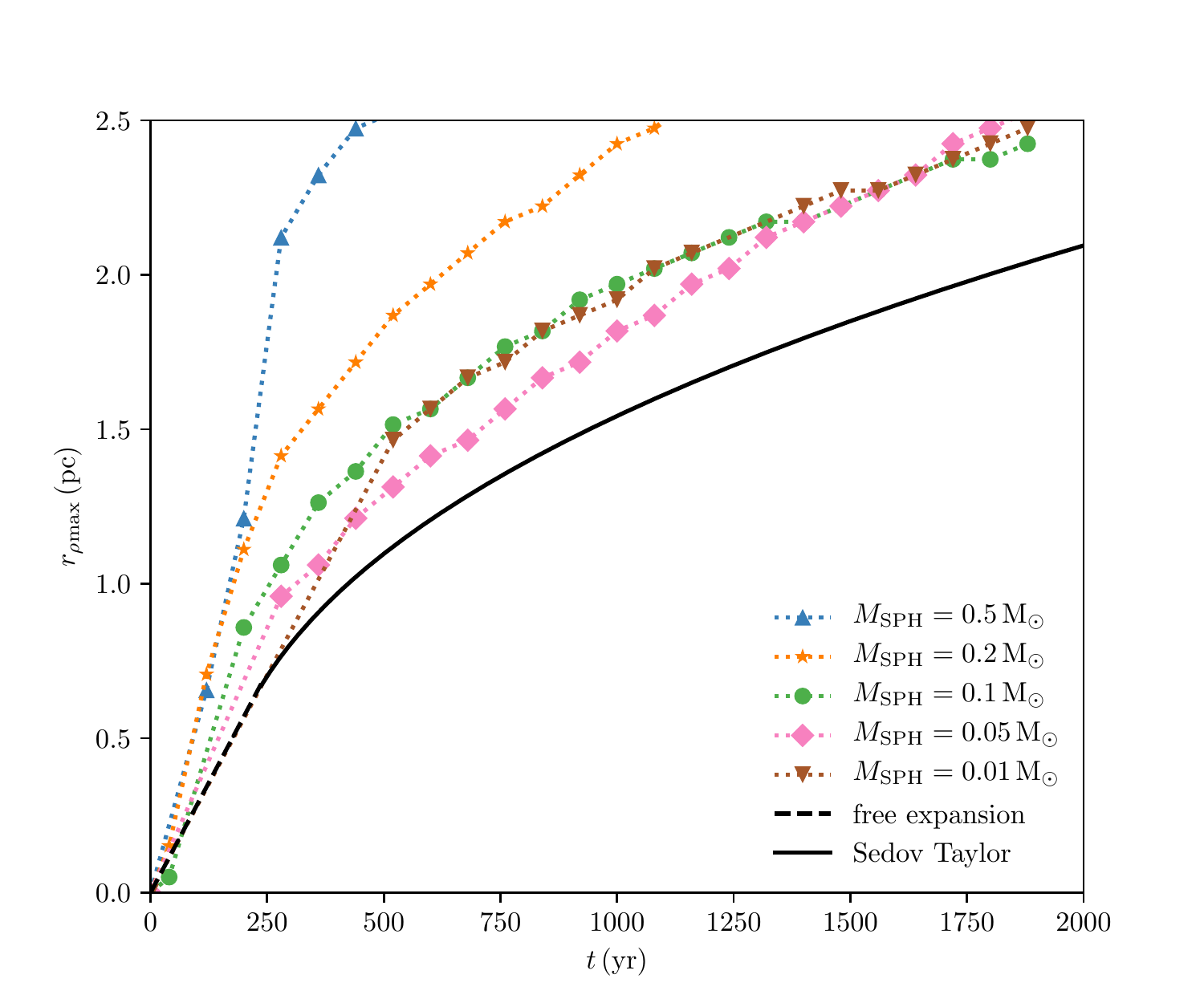}
            \caption{
                The radius with the highest mean gas density ($r_{\rho \mathrm{max}}$) as a function of time for the supernova test.
                We include the analytical solution, which is free expansion followed by the Sedov-Taylor solution.
                \label{fig: sn radius time}}
            \end{figure}

        % supernova test, same as em but with supernova
        As discussed in Section~\ref{sec: heating}, the heating wind mode can also be used to simulate the effect of a supernova on the surrounding gas.
        The supernova test (Table~\ref{tab: sn parameters}) is similar to the embedded star test (Section \ref{sec: em}).
        We start the simulation by evolving the star using the stellar evolution code \seba\ \citep{portegieszwart_seba_2012} to a few timesteps ($\unsim40$~yr) before the star goes supernova ($\unsim9.78$~Myr).
        We place the star inside the uniform density gas medium and use the option to derive the stellar wind parameters from the output of a stellar evolution calculation (see Section~\ref{sec: wind from stev}).
        When this option is set, \stellarwind\ detects the supernova and creates the particles with a combined mass of 12.95~\MSun\ and an energy of $10^{51}$~erg.
        After the supernova feedback is generated we trace the resulting blast wave.
        Similar to the embedded star test, we get a sphere of high density material moving away from the star, however, due to the higher energy input, the radial velocity is higher.
        Using test simulations, we have found that a timestep of $t_\mathrm{wind} = 20$~yr is required to avoid time-stepping artefacts with this high velocity \citep[also see:][]{pelupessy_evolution_2012}.

        % In figure ... top, the higher density in the center is a wave
        In Figure~\ref{fig: sn dens profile} we present the radial mean density profile for simulations with four different particle masses.
        We find that for the low resolution simulation ($M_\mathrm{SPH} = 1$~\MSun), the gas has moved away from the star, but the shape of the shockfront, where the density is highest, is only loosely defined.
        For the higher resolution simulations, we do see the shape of the main shockfront clearly, and the simulations agree on the radius of highest density at $t = 1000$~yr.
        However, the radius of the shockfront does not converge to the analytical solution.
        There is an increased density at $r = 0$ for the $M_\mathrm{SPH} = 0.01$~\MSun\ simulation.
        This feature is present at some point for all high resolution simulations and is the result of a reverse density wave within the outgoing shockwave.
        These waves are an artefact of the hydrodynamical simulation method used, but they are not an accurate representation of the true physical process.
        They should not be confused with the reverse shock that takes place in real supernova remnants.
        While these density waves do subside after a few thousand years, the density inside the supernova bubble shortly after the explosion should not be considered correct.

        % analytical solutions
        The time evolution of the expansion of gas from a supernova explosion is usually modelled in separate phases.
        The first phase is a free expansion, where the ejected gas moves at an approximately constant velocity, sweeping up the gas in the interstellar medium.
        After the mass of swept up gas is equal to the mass of the originally expelled gas, the expansion can be approximated as a pure adiabatic expansion, which is described by the self-similar Sedov-Taylor solution.
        Only this last phase can be simulated with the heating wind mode of \stellarwind, because the particles are given a high internal energy instead of an initial velocity.

        % In figure ... bottom
        In Figure~\ref{fig: sn radius time} we present the time evolution of $r_{\rho \mathrm{max}}$, which we calculate in the same way as for the embedded star test.
        We now compare it to the analytical solution for the two phases of a supernova blast wave, the free expansion and the Sedov-Taylor solution.
        We find that the simulations do approach the analytical solution and roughly follow the same shape, but even the highest resolution simulation expands faster than the analytical solution.
        We do not model the initial free expansion phase and shockwaves are in general difficult to simulate using SPH \citep[e.g.][]{hubber_convergence_2013}.
        Differences with the analytical solution are therefore to be expected and this type of simulation should be interpreted with care.

        % But we do have something going for these simulations
        Given these caveats, both the use of SPH and the chosen approximations may not seem to be the ideal choice for simulating a supernova explosion in a gaseous medium.
        Indeed, depending on the goal of the simulations, other available methods could be more suitable, for example using a grid based simulation code \citep[e.g.][]{rogers_feedback_2013} or including magnetic fields \citep[e.g.][]{kortgen_supernova_2016}.
        However, the method presented here has two main advantages:
        1) It is simple and scales well to very low SPH resolution, making it computationally faster than more detailed simulation techniques.
        2) The use of SPH combined with \bridge\ allows easy gravitational coupling between the gas and the stars.
        We can therefore use this code to run large scale simulations of multiple supernova explosions in a gaseous medium also containing many dynamic stars.
        These advantages allow us to model a very turbulent stage in the evolution of embedded star clusters.

    \subsection{Colliding wind triple}
        \label{sec: tr}

        \begin{table}
            \caption{
                The stellar, wind and orbital parameters of the colliding wind triple simulation.
                \label{tab: tr parameters}}
	       \begin{adjustbox}{width=0.48\textwidth}
            \begin{tabular}{l l l l l}
                \B name & parameter & star 1 & star 2 & star 3 \\
                \hline
                \T stellar type & & WR5 & O6 & O9.5 Giant \\
                mass-loss rate & $\Mdot$ (\MSun/yr) & $1.8 \cdot 10^{-5}$ & $10^{-6}$ & $10^{-6}$ \\
                wind velocity & $\vinf$ (km/s) & 2000 & 1000 & 1000 \\
                mass & $M_*$ (\MSun) & 12 & 20 & 30 \\
                radius & $R_*$ (\RSun) & 2.2 & 10 & 50 \\
                luminosity & $L_*$ (\LSun) & $2 \cdot 10^5$ & $1.4 \cdot 10^5$ & $5.5 \cdot 10^4$ \\
                temperature & $T_*$ (K) & $7.1 \cdot 10^4$ & $4.5 \cdot 10^4$ & $3.9 \cdot 10^4$ \\ [2mm]
                orbital period & $p$ & \multicolumn{2}{c}{19.14~days} & 130~yr \\
                eccentricity & $e$ & \multicolumn{2}{c}{0} & 0 \\
                \B inclination & $i$ & & & $0^{\circ}$ \\
                \hline
                \T wind timestep & $t_\mathrm{wind}$ & 0.2~days  & & \\
                end time & $t_\mathrm{end}$ & 190~days & & \\
                particle mass & $M_\mathrm{SPH}$ & $10^{-11}$~\MSun & & \\
                \end{tabular}
                \end{adjustbox}
            \end{table}

        \begin{figure*}[!tbp]
            \centering
            \includegraphics[width=0.88\textwidth]{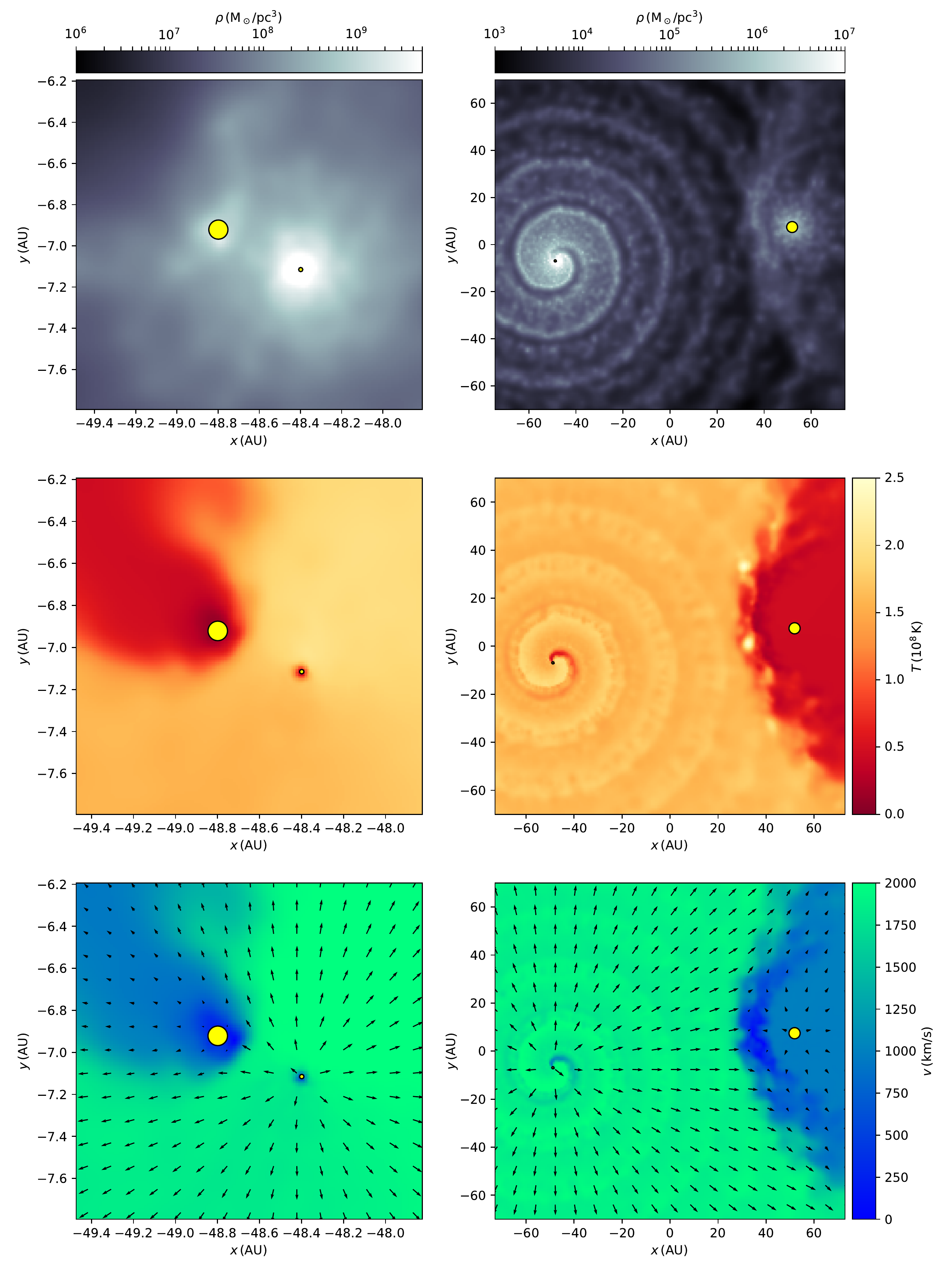}
            \caption{
                The gas density (top), temperature (center) and velocity (bottom) in the orbital plane of the colliding wind triple simulation for the inner (left) and outer (right) binary.
                The sizes of the stars (yellow circles) in the plots on the right hand panels were multiplied by 10 to make them visible.
                In the left hand panels, the Wolf-Rayet star (star~1, see Table~\ref{tab: tr parameters}) can be seen on the right and star~2 on the left.
                In the right hand panels, the short period binary (star 1 and 2), can be seen on the left and the O9.5 supergiant (star~3) on the right.
                In the bottom plots, the arrows indicate the wind direction and larger arrows correspond to higher wind velocities, however, the colors provide a more precise indication of the velocities.
                Note that the two density plots have separate color bars, while the temperature and velocity plots each share a single color bar.
                \label{fig: tr heat}}
            \end{figure*}

        % Real example, colliding wind triple WR48
        The previous tests were for single stars and therefore the geometry of the outflow was not modified by the environment.
        In this test (Table~\ref{tab: tr parameters}), we simulate a triple star system where all three stars have a strong stellar wind.
        The system we simulate is loosely based\footnote{
            We are aware that the chosen values do not match the most up-to-date observations of the WR48 system.
            However, the goal of this investigation is to demonstrate the use of the \stellarwind\ code, not to explain the observed system.}
        on WR48 ($\theta$~Muscae), which is a triple system \citep{sugawara_redshifted_2008} consisting of a WC5/WC6 + O6/O7V binary with a short period \citep[$\unsim 19$~days,][]{hill_modelling_2002} and an O9.5/B0Iab star in a longer orbit \citep[$>130$~yrs,][]{dougherty_non-thermal_2000} around that binary.
        For this simulation we have used similar numerical parameters to the fast wind test in Section~\ref{sec: fw}.

        % We get some very cool effects
        In Figure~\ref{fig: tr heat} we show the gas density, temperature and velocity at the end of the simulation.
        Due to the large difference between the inner and outer orbital periods, the system appears similar to a normal colliding wind binary, which was assumed in previous models of WR48 \citep{hill_modelling_2002}.
        However, the orbital motion of the inner binary creates a spiral pattern in the density and temperature distribution, which is very different from the wind from a single star.
        This spiral pattern creates high and low temperature regions in the shockfront where the wind from the inner binary collides with the wind from the third star.
        The observations of this shockfront can therefore be quite different from observations of a normal colliding wind binary.

        % This is just an example to show cool simulations are possible
        It is important to note that this simulation is just an example of what is possible with the \stellarwind\ module and not an in-depth investigation into wind interactions in a triple star systems.
        For example, in the middle panels of Figure~\ref{fig: tr heat} we can see that the temperature of the wind from the inner binary is extremely high ($>10^8$~K).
        These high temperatures are unrealistic because in reality the gas would cool, which is not taken into account in this simulation.
        When using this code for simulations that are to be compared with observations, gas cooling and a convergence test of the shockfront regions should be performed.

\section{Discussion and Conclusion}
        \label{sec: Discussion and Conclusion}

        \begin{table*}
            \caption{
                An overview of the modes in \stellarwind\ and their suggested application domains.
                \label{tab: modes}}

            \begin{tabular}{l l p{7cm} p{7cm}}
                \B mode & section & description & application \\
                \hline
                \T simple & \ref{sec: simple}
                    & Creates particles with a radial velocity given by the desired terminal wind velocity.
                    & Detailed wind interaction simulations well outside the acceleration zone and past the sonic point.\\[2mm]
                accelerating
                    & \ref{sec: accelerating} & Similar to simple wind, but also accelerate particles near the star.
                    & Detailed wind simulation near or inside the acceleration zone and near the sonic point. \\[2mm]
                heating & \ref{sec: heating}
                    & Does not give new particles a radial velocity, but instead adds internal energy to the particles.
                    & Large scale, low resolution simulations of wind from embedded stars, including the effect of a supernova. \\
                \end{tabular}
            \end{table*}

        % We presented and tested the code
        We have presented and tested the \stellarwind\ module, which can be used to simulate stellar winds within the \amuse\ framework by creating (and accelerating) SPH particles.
        The code includes three different modes: the simple mode, the accelerating mode  and the heating mode (Table~\ref{tab: modes}).
        We have tested the code for single stars with fast and slow winds, as well as an embedded star with both wind and a supernova explosion.
        For both fast and slow winds, the simple and accelerating wind modes perform well, although subsonic winds must be simulated with the latter.
        For the embedded star, the heating wind mode creates a wind bubble, even at low resolution; with higher resolution the expansion profile approaches the analytical solution.
        After a supernova, the heating wind mode creates an expanding shell with velocities similar to the analytical solution if small enough timesteps are used.
        Finally we have shown an example of how this module can be used to tackle new problems, by simulating a colliding wind triple system.

        % Projects that are under way
        The \stellarwind\ module can be used for many different simulations that involve stellar winds and several projects are already in progress.
        The simple wind mode has been used to simulate the accretion of gas from the winds of the S-stars onto the super-massive black hole Sgr~A$^*$ \citep{leutzgendorf_stellar_2016}.
        The accelerating wind mode is used to simulate the accretion of the wind from a red giant onto a close binary companion \citep{saladino_2018}.
        The heating wind mode is part of a larger simulation to investigate the evolution of the Arches cluster \inprep{van der Helm et al.}.
        In Table~\ref{tab: modes} we give an overview of the application of the different modes.
        The code is publicly available in the \amuse\ framework.

        % What can be added to the code and AMUSE
        There are many other types of simulations involving stellar winds that could be done with the \amuse\ framework and corresponding modes could be added to \stellarwind.
        It would be possible to add the mass and corresponding energy lost by stars to existing SPH particles.
        This would make it possible to run simulations of embedded stars with even lower resolution (higher SPH particle mass).
        However, this would result in unequal mass particles, which requires advanced treatment in the SPH codes.
        In the other extreme, since radiative transfer codes are available in \amuse, it would be possible to add a mode that solves the radiative hydrodynamics of the wind and this would make detailed stellar wind simulations possible.
        In fact, such coupled simulations have been performed with \amuse\ already \citep[][N. Clementel, private communication]{wall_simulating_2017}.

\begin{acknowledgements}
    We thank N. L{\"u}tzgendorf for testing and improving the simple wind mode, R. P. Kudritzki for his advice on the \vinf\ scaling law and F. I. Pelupessi for providing his code for the heating wind mode.
    This work was supported by the Netherlands Research Council NWO.
    \end{acknowledgements}

% bibliography
    \bibliographystyle{aa}
    \bibliography{wind_paper_refs}

\appendix
\section{Equations}
        \label{sec: equations}

        In this appendix, we calculate the analytical predictions for a stationary, spherically symmetric wind which are used in \stellarwind.
        For these calculations, we assume that the mass-loss rate (\Mdot) and the velocity as a function of radius ($v(r)$) are known and we define the acceleration
        \begin{equation}
            \label{eq: acceleration}
            a(r) = \frac{dv}{dt} = \frac{dv}{dr} \frac{dr}{dt} = v(r) \frac{dv}{dr}.
            \end{equation}

    \subsection{Radius as a function of time}
        \label{sec: radius at time}
        To calculate the outer radius of a new wind shell, we need to know the radius as a function of time ($r(t)$) where the wind starts at the stellar surface, so $r(0) = R_*$.
        Since $v(r)$ is known, we can write
        \begin{equation}
            \begin{split}
            v(r(t)) & = \frac{d r(t)}{dt}, \\
            dt & = \frac{dr}{v(r)},
            \end{split}
            \end{equation}
        which is solved by,
        \begin{equation}
            t = \int_{R_*}^{r(t)} \frac{1}{v(r)} \, dr.
            \end{equation}
        In general this equation has to be solved numerically\footnote{
            When solving the equations mentioned here numerically, we use the python library \textsc{scipy} (scipy.org).
            For integrals we use \textsc{scipy.integrate.quad} and for finding a root we use \textsc{scipy.optimize.brentq}.
            See docs.scipy.org for the details of these methods.
        } for $r(t)$, although for some velocity functions we can solve it analytically, for example if $v(r) = \vinf$ then
        \begin{equation}
            \begin{split}
            t & = \frac{1}{\vinf} \int_{R_*}^{r(t)} dr = \frac{r(t) - R_*}{\vinf}, \\
            r(t) & = R_* + t * \vinf.
            \end{split}
            \end{equation}

    \subsection{New particle radii}
        \label{sec: new particle radii}
        When we create a new shell of particles, we want the density profile in the shell to match the density profile corresponding to the chosen velocity profile.
        To calculate that density profile, we first note that the mass-loss rate, $\dot{M}$ is related to the density and the velocity at any point of the wind via the equation of mass continuity,
        \begin{equation}
            \label{eq: continuity}
            \Mdot = 4\pi r^2 \rho(r) v(r),
            \end{equation}
        where $\rho$ is the density of the wind.
        Because we assume that \Mdot\ and $v(r)$ are known, we can rewrite this as
        \begin{equation}
            \label{eq: density}
            \rho(r) = \frac{\Mdot}{4\pi r^2 v(r)}.
            \end{equation}

        To generate the positions of new particles, we start with a cube filled with particle positions with a uniform density.
        In our code, this can be a simple grid or randomly generated positions.
        From that cube, we remove all particles that are not inside the desired shell to get a shell of particles with uniform density.
        After that, we shift the particle positions in the radial direction to get the desired density profile.

        To find the new particle radius, we define the relative enclosed mass, $x$ as
        \begin{equation}
            \label{eq: encl mass}
            x = \frac{\int_{R_*}^{r_p} \pi r^2 \rho(r) dr}{\int_{R_*}^{r(t)} \pi r^2 \rho(r) dr},
            \end{equation}
        where $r_p$ is the radius of the particle and $R_*$ and $r(t)$ are the inner and outer radius of the shell respectively.
        For the uniform density shell that was generated, this reduces to
        \begin{equation}
            x_u = \frac{\int_{R_*}^{r_p} r^2 dr}{\int_{R_*}^{r(t)} r^2 dr} = \frac{r_p^3 - R_*^3}{r(t)^3 - R_*^3}.
            \end{equation}
        For the desired density profile based on a given velocity profile, we rewrite equation~\ref{eq: encl mass} in terms of $v$ using equation~\ref{eq: density}
        \begin{equation}
            \label{eq: x of v}
            x_v = \frac{\int_{R_*}^{r_p} \frac{\Mdot}{v(r)} dr}{\int_{R_*}^{r(t)} \frac{\Mdot}{v(r)} dr} = \frac{\int_{R_*}^{r_p} \frac{1}{v(r)} dr}{\int_{R_*}^{r(t)} \frac{1}{v(r)} dr}.
            \end{equation}
        We then set $x_u = x_v$ where $x_u$ is calculated with the old particle radius (of the generated uniform density shell).
        The last step is to solve equation~\ref{eq: x of v} to get the new particle radius $r_p$.
        In general this equation has to be solved numerically, although for some velocity functions we can solve it analytically, for example if $v(r) = \vinf$ then
        \begin{equation}
            \begin{split}
            x & = \frac{\int_{R_*}^{r_p} \frac{1}{\vinf} dr}{\int_{R_*}^{r(t)} \frac{1}{\vinf} dr} = \frac{\int_{R_*}^{r_p} dr}{\int_{R_*}^{r(t)} dr} = \frac{r_p - R_*}{r(t) - R_*}, \\
            r_p & = R_* + x (r(t) - R_*).
            \end{split}
            \end{equation}

    \subsection{Gas pressure}
        \label{sec: gas pressure}
        To calculate the expected acceleration, $a_P(r)$ caused by the gradient of the gas pressure, $P(r)$ we assume a polytropic equation of state,
        \begin{equation}
            \label{eq: polytropic eos}
            P = K \rho(r)^{\gamma},
            \end{equation}
        where $K$ is the polytropic constant and $\gamma = 5/3$ is the adiabatic index for a monoatomic ideal gas.
        Because $K$ is constant we can calculate it at the surface of the star and use that value for the entire wind.
        To calculate $P(R_*)$ we use
        \begin{equation}
            \label{eq: pressure}
            P(r) = (\gamma - 1)\rho(r) u,
            \end{equation}
        where $u$ is the internal energy of the gas particles defined by
        \begin{equation}
            u = \frac{3}{2} \frac{k_B T_*}{\mu},
            \end{equation}
        where $k_B$ is the Boltzmann constant, $T_*$ is the temperature at the photosphere of the star and $\mu$ is the mean molecular weight of the gas particles.
        Combining equations~\ref{eq: polytropic eos}~and~\ref{eq: pressure} we get
        \begin{equation}
            K = u(\gamma-1)\rho(R_*)^{1-\gamma}.
            \end{equation}

        The acceleration caused by the gradient of the gas pressure is
        \begin{equation}
            a_P(r) = - \frac{1}{\rho(r)} \frac{\partial P(r)}{\partial r},
            \end{equation}
        which we can rewrite using equations~\ref{eq: polytropic eos}~and~\ref{eq: density}
        \begin{equation}
            \begin{split}
            a_P(r) & = - \frac{K}{\rho(r)} \frac{\partial \rho^{\gamma}}{\partial r} \\
            & = - \frac{K}{\rho(r)} \gamma \rho(r)^{\gamma - 1} \frac{\partial }{\partial r} \frac{\Mdot}{4\pi r^2 v(r)} \\
            & = K \gamma \rho(r)^{\gamma - 1} \left( \frac{2}{r} + \frac{1}{v(r)} \frac{dv(r)}{dr} \right ).
            \end{split}
            \end{equation}

    \subsection{Density as a function of radius}
        \label{sec: dens radius}
        In Sections~\ref{sec: em}~and~\ref{sec: sn} we calculate the density as a function of radius.
        For each radius $r$, we take six points in six directions ($+r$ and $-r$ along each axis x, y and z) and calculate the SPH density at those points.
        Note that there does not need to be an SPH particle at that point to calculate the density.
        We then take the mean of these 6 densities to be the density at that radius.
        To calculate the radius with maximum density $r_{\rho \mathrm{max}}$, we calculate this for a grid of radii and select the radius with the largest density.

\end{document}